# *Hydra* morphogenesis as phase-transition dynamics

Oded Agam[1] and Erez Braun[2]

[1]*The Racah Institute of Physics, Edmond J. Safra Campus, The Hebrew University of Jerusalem, Jerusalem 9190401, Israel.*
[2] *Department of Physics and Network Biology Research Laboratories, Technion-Israel Institute of Technology, Haifa 32000, Israel.*

**ABSTRACT**

We utilize whole-body *Hydra* regeneration from a small tissue segment to develop a physics framework for animal morphogenesis. Introducing experimental controls over this process, an external electric field and a drug that blocks gap junctions, allows us to characterize the essential step in the morphological transition - from a spherical shape to an elongated spheroid. We find that spatial fluctuations of the $Ca^{2+}$ distribution in the *Hydra's* tissue drive this transition and construct a field-theoretic model that explains the morphological transition as a first-order-like phase transition resulting from the coupling of the $Ca^{2+}$ field and the tissue's local curvature. Various predictions of this model are verified experimentally.

How does a robust stereotypical body-form emerge in animal development? Recent works show that morphogenesis, at different stages of development and in a variety of organisms, involves multiple interconnected biochemical[1, 2], mechanical[3-11], and electrical[12-16] processes. From a physical perspective, morphogenesis is far-from-equilibrium pattern formation dynamics driven by active internal forces. Despite recent progress, a framework describing the coordination and integration of the processes underlying morphological transition in morphogenesis remains a major challenge. In particular, it is unclear what are the primary fields that drive the morphological transition and whether morphogenesis is a slow incremental process reflecting the accumulation of small changes, a sharp transition, or an avalanche of a series of transitions. This Letter aims as a step toward gaining insight into these fundamental issues.

*Hydra*, a small freshwater multicellular organism made out, essentially, of a bilayer epithelial tissue with a uniaxial geometry[17-19], provides a unique system for such an investigation[7, 20-25]. Whole-body *Hydra* can be regenerated from a small tissue segment almost without cell division[23, 26], see Fig. 1a and movies 1-2[27]. An excised tissue fragment first forms a closed spheroidal shape and then elongates towards a cylindrical shape - the body-form of a mature *Hydra*[7, 21]. The appearance of tentacles, a head and a foot, completes the developmental process.

An essential experimental step in understanding the nature of the morphological transition is to introduce controls that affect the pattern formation process. Recently, one of us showed that *Hydra* morphogenesis could be modulated on demand; regeneration can be halted in a reversible way and even be reversed when the tissue is subjected to an external electric field[13]. *Heptanol* - a drug that blocks gap junctions responsible for the long-range electrical communication in the tissue - was also shown by us to reversibly halt the morphological transition in *Hydra* regeneration[12].

An external electric field and *Heptanol* strongly affect the calcium ($Ca^{2+}$) activity in the *Hydra*'s epithelial tissue. While the external electric field enhances the $Ca^{2+}$ activity and its spatial correlations, blocking gap-junction suppresses them[12]. The *Hydra* epithelial tissue is a muscle, and the $Ca^{2+}$ activity controls its internal contractile forces by enabling the operation of myosin motors on the actin fibers[20, 28-32]. The mechanical balance between these forces and the pressure applied by the fluid within the cavity, enclosed by the epithelial bilayer tissue, determines the tissue's morphology[7, 22, 24, 33-35]. Thus, the $Ca^{2+}$ activity directly affects the shape of the *Hydra*'s tissue, making the external electric field and *Heptanol* effective experimental controls of its morphology.

In a previous study, we showed that the experimental probability distributions of the spatial configurations of the $Ca^{2+}$ field, $\phi$, as well as their dependence on the controls, can be well modeled by $P[\phi] = \exp(-S_{\text{Ca}})/Z_{\text{Ca}}$, where $Z_{\text{Ca}}$ is a normalization constant, and

$$S_{\text{Ca}} = \oiint d^2x\sqrt{g}\left[\frac{D}{2}g^{\mu\nu}\left(\partial_\mu\phi\right)\left(\partial_\nu\phi\right) + U\left(\phi\right)\right] . \qquad (1)$$

Here, $U(\phi)$ is a tilted double-well potential reflecting the electrically excitable nature of the *Hydra*'s epithelial tissue, while $D$ is a stiffness parameter that limits large spatial gradients of $\phi$. The integral in Eq. (1) is over the whole *Hydra* closed surface, parametrized by a two-component vector $\boldsymbol{x}$, and $g$ is the determinant of the corresponding metric tensor $g_{\mu\nu}$. This model ignores the coupling of the $Ca^{2+}$ field to the tissue's morphological structure.

Here, we experimentally characterize the significant morphological transition of the regenerating *Hydra*'s tissue, from a spheroidal shape into a persistent elongated cylindrical shape (red arrow of Fig. 1a) and show that a simplified field-theoretic model, based on the coupling between the $Ca^{2+}$ field and the scalar (Gaussian) curvature of the tissue, captures the main features of the experimental data. In contrast with other models of epithelial morphogenesis, which describe the system at a cell resolution level (e.g., vertex models[36, 37]), our model is a coarse-grained description of whole-tissue morphogenesis. In particular, it shows that the morphological transition resembles the dynamics of a phase transition of the first-order type. It also explains the mechanisms by which the *Heptanol* and the electric field halt regeneration and provides three insights that we verified experimentally: (a) Negative correlations between the fluctuations in the scalar curvature of the tissue and the local $Ca^{2+}$ activity. (b) Reversal of a mature *Hydra*'s morphology from a cylindrical shape back into a spheroidal one by weakening the internal muscle forces by *Heptanol*, and (c) re-initiation of the morphogenesis process by the application of an electric field simultaneously with *Heptanol* that halts regeneration.

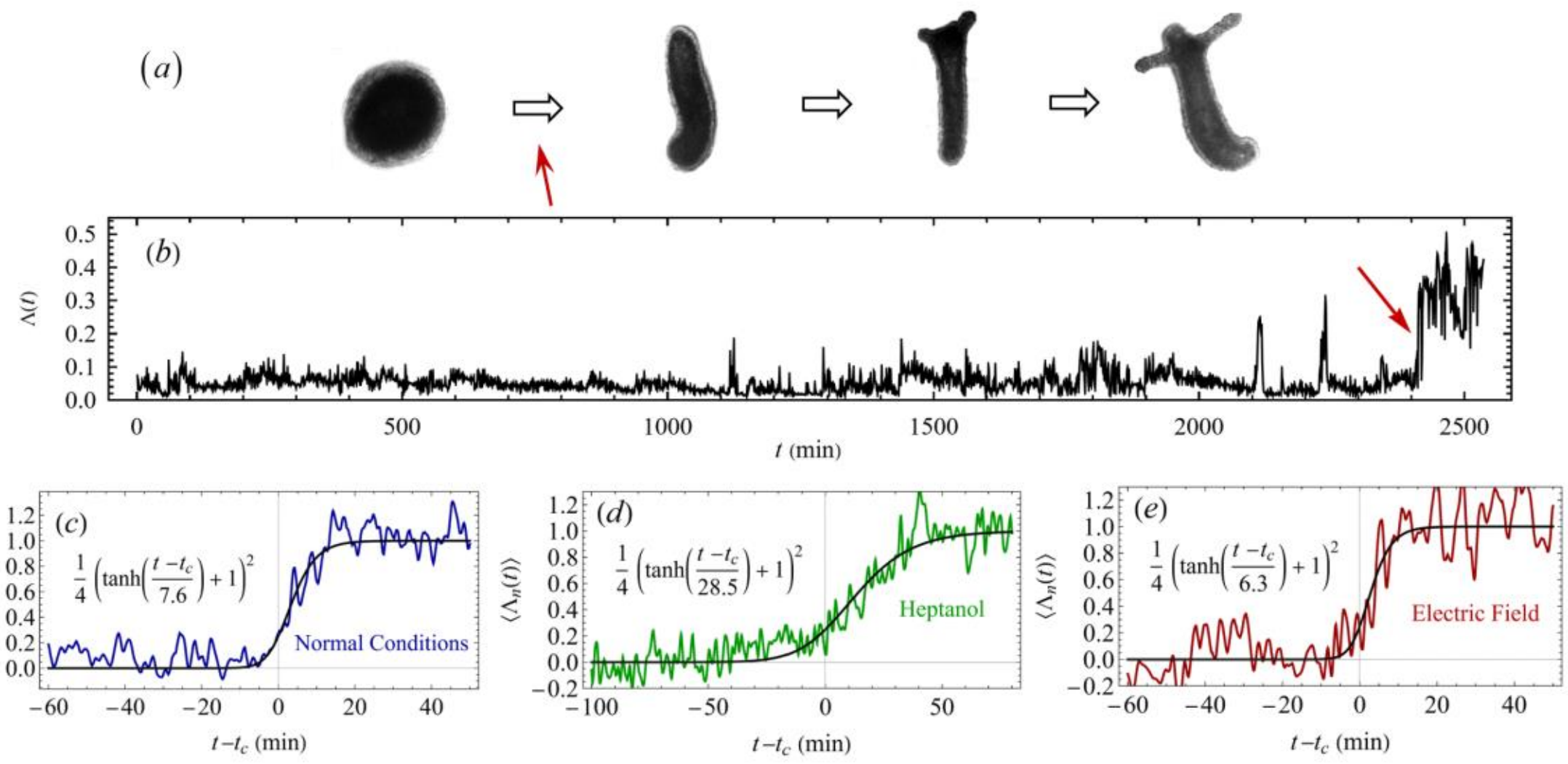

**FIG. 1: Morphogenesis as phase transition dynamics.** (a) A stereotypical regeneration of the *Hydra*'s tissue from a spheroidal (approximately 300 μm in diameter) to a cylindrical shape (red arrow), complemented by the appearance of tentacles in the mature, fully developed animal. (b) An example time trace of the shape parameter $\Lambda$ that characterizes the morphology of the *Hydra*'s tissue under normal conditions ( $t=0$ is the start of the experimental recording of the spheroidal closed tissue segment). This parameter is zero when the tissue is spherical and approaches one when it becomes an elongated cylinder. Most of the time, the tissue fluctuates near a spherical shape. The transition, marked by the red arrow, occurs within a few minutes. The lower panels show time traces of the normalized shape parameter $\Lambda_n$ around the transition time $t_c$ , averaged over tissue samples: (c) seven tissue samples for normal regeneration in a normal *Hydra* medium (HM) , (d) six tissue samples regenerating under *Heptanol* (300 μl/l in HM), and (e) six tissue samples regenerating under an external electric field (30V/4mm in HM) (e). The experimental data was fitted to a smooth step function, $a+b\tanh\left[\left(t-c\right)/d\right]$ and then normalized by shifting the transition point to zero, subtracting $a-b$ , and dividing by $2b$ . Thus, the normalized data features a transition between $\Lambda_n=0$ and $\Lambda_n=1$ , which occurs at $t=0$ . Finally, the normalized data of each sample is represented by an interpolating function, and the resulting functions are averaged over the ensemble of samples. The solid black curves represent the instantons that describe the transitions ( $q^2\left(t\right)$ from Eq. 8).

In the experiment, tissue fragments are excised from the middle regions of mature *Hydra*, expressing a fast $Ca^{2+}$ fluorescence probe (GCaMP6s) in its epithelial (endoderm) cells[12, 13, 32]. The fragments were allowed to fold into spheroids for ~3 hrs and then placed in the experimental setup under a fluorescence microscope (see[12, 27] for details of the experimental methods).

To characterize the morphological transition, we define the tissue's shape parameter $\Lambda=1-4\pi A/L^2$ , where $A$ is the tissue's projected area and $L$ is its perimeter. For a spherical tissue $\Lambda=0$, while any deformation increases its value. This shape parameter faithfully characterizes the *Hydra*'s tissue form, as long as its regeneration is far from completion. At this period, the tissue is approximately a spheroid whose axis is parallel to the microscopy imaging projection plane. Shape fluctuations of the spheroidal tissue towards a prolate shape occur primarily along the body-axis of the regenerating *Hydra*, pre-determined by the ectoderm longitudinal supracellular actin fibers[7, 20]. These shape fluctuations, together with gravity and friction, predominantly align the spheroid axis parallel to the imaging projection plane.

Fig. 1b shows a typical time trace of $\Lambda\left(t\right)$ for a *Hydra*'s tissue regenerating under normal conditions. The transition from a spheroidal shape into a cylindrical one (red arrow in Fig. 1b) occurs within a strikingly short time period. A zoom over the normalized shape parameter around the transition region, averaged over six tissue samples, is depicted in Fig. 1c. It shows that the transition occurs over approximately eight minutes. This is an extremely short time scale compared to the typical mechanical relaxation time scale of the tissue, measured to be around 100 minutes[12], and following a long duration of many hours in which the *Hydra*'s tissue remains approximately spherical.

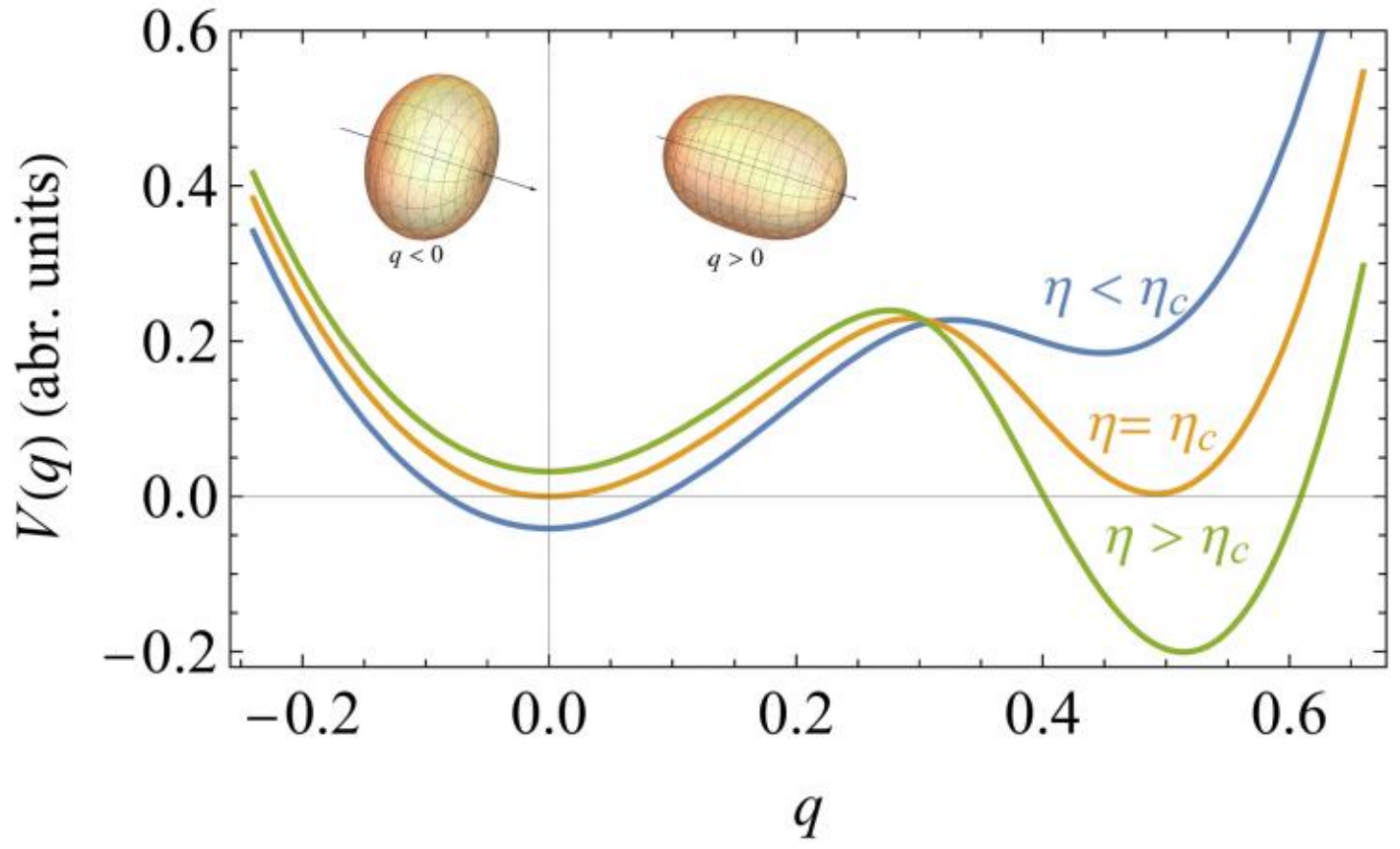

**FIG. 2: The morphological potential.** The potential $V(q)$ for several values of the coupling strength between the $Ca^{2+}$ field and the tissue's curvature $\eta$. The minimum at $q=0$ represents a spherical tissue shape, while that at $q>0$ is associated with an approximate prolate spheroidal shape, right after the first stage of the morphological transition. Negative and positive values of $q$ are associated with oblate and prolate ellipsoids, respectively, as demonstrated by the insets. The second minimum becomes a global minimum when the coupling parameter between the $Ca^{2+}$ field and the tissue's curvature exceeds a critical value, $\eta_c$. The morphological evolution is the dynamical process of a transition from the metastable spherical state to the lower energy elongated state, described by Eq. (7).

*Heptanol* allows for halting the morphological transition, but variability among different tissue samples leads some of them to escape and proceed along the regeneration trajectory (~20-30% of the samples). Again, these samples show a sharp morphological transition but more moderate than the transition under normal conditions, see Fig. 1d. Here, the typical transition time is 28 minutes.

Similarly, slightly below the critical electric field that halts regeneration[12, 13], tissue samples proceed along the regeneration trajectory. These samples also show a sharp morphological transition, as demonstrated in Fig. 1e.

Examples of area and shape parameter traces for individual tissue samples under the different controls are shown in the supplementary material (SM)[27].

To understand these results, we consider a simplified model of the *Hydra*'s tissue, ignoring its bilayer structure and regarding the tissue as a closed 2D surface[38] with a minimal coupling of its curvature to the $Ca^{2+}$ field, $\phi$, given by the Jackiw–Teitelboim action [39, 40]:

$$S_{\text{int}} = \eta \oiint d^2x\sqrt{g}\,\phi\mathcal{R}\,. \qquad (2)$$

Here $\mathcal{R}$ is the scalar curvature and $\eta$ is the coupling strength. While the mechanism of the effect of $Ca^{2+}$ on the curvature is understood (see SM[27]), the above coupling action also implies a feedback effect of curvature on the calcium activity. The feedback mechanism is poorly understood but should exist for effective control of morphogenesis[3, 41-44]. It is also supported by the Granger causality analysis provided in the SM[27].

The third component of our model is the action that dictates the probability of the tissue's morphological configurations. The *Hydra* tissue is a muscle made of cells that actively stretch and compress. It cannot be considered an elastic sheet or a simple membrane. The obviation that its area may change dramatically near the morphological transition (see Figs. S1-S3 in the SM[27]) implies that morphological changes which do not preserve isometry are very soft. At the same time, the distinctive manner by which an excised, almost flat, tissue fragment encloses on itself to form a spheroidal shape, suggests that the tissue curvature plays a central role in determining its morphology. Taking into account that the scalar curvature of a closed surface is a topological invariant, the simplest action that governs the tissue shape is

$$S_K = B\oiint d^2x\sqrt{g}\left(\mathcal{K}-\mathcal{K}_0\right)^2, \qquad (3)$$

where $\mathcal{K}$ is the mean curvature, and $\mathcal{K}_0$ is the spontaneous curvature of the tissue in the absence of constraints. This action has the same form of the Canham-Helfrich free energy of membranes[45, 46] where $B$ is analogous to the bending modulus. However, its microscopic origin is not well understood.

Our modeling starting point is the total action,

$$S = S_{\text{Ca}} + S_{\text{int}} + S_K\,, \qquad (4)$$

which, together with a constraint over the volume enclosed by the tissue, dictates the probability, $\exp(-S)/Z$ ($Z$ is the normalization constant), of the various configurations of the system. For justification of this model and discussion of its extensions, we refer the reader to the SM[27].

Consider axisymmetric deformations of the tissue's spherical state described by a surface of revolution. Its distance from the origin as a function of the polar coordinate, $\theta$, is given by

$$R(\theta) = R_0(q)\left[1 + q\left(3\cos^2\theta - 1\right)\right] \quad , \tag{5}$$

where $-0.5 < q < 1$ to prevent self-intersections of the surface and $R_0(q)$ is a positive function, chosen such that the volume enclosed by the tissue is fixed. The angular dependence in Eq. (5) is proportional to the spherical harmonic function $Y_2^0(\theta)$, which is the leading term describing a symmetric uniaxial deformation. Depending on the sign of the parameter $q$, $R(\theta)$ describes a prolate or an oblate surface of revolution, as shown in the insets of Fig. 2. Substituting (5) in (4) and integrating over the configurations of the $Ca^{2+}$ field yields the potential $V(q)$ that determines the possible stable and metastable states of the system:

$$\exp\left[-V(q)\right] = \frac{1}{Z}\int \mathcal{D}\phi \exp(-S). \tag{6}$$

The evaluation of this integral, assuming the metastable states of the system are long-lived, yields the potential plotted in Fig. 2 (for details of the calculation, see SM[27]). This potential has two minima: One at $q = 0$ describing a spherical tissue and another one at $q = q_0 > 0$ describing a prolate shape. For sufficiently strong coupling, $\eta > \eta_c$, the spherical state becomes unstable. Thus, a tissue prepared in the spherical state will transform into an elongated shape by an activation over the barrier. This transition can be approximately described by the *Langevin* equation:

$$\frac{dq}{dt} = -\frac{d}{dq}V(q) + \xi(t), \tag{7}$$

where $\xi(t)$ is a stationary Gaussian white noise instantaneous correlations: $\langle \xi(t)\xi(t')\rangle = \sigma^2\delta(t-t')$, and zero mean. The source of this noise is the fast temporal fluctuations of the $Ca^{2+}$ field.

To characterize the solution of Eq. (7), we approximate the double well potential by $V(q) \simeq 16\lambda q^2(q-q_0)^2 + \mu q$ where $\lambda$ controls the barrier height and $\mu$ determines the tilt. For $|\mu| \ll \lambda$ and an intermediate level of the noise strength, the instanton that describes the transition[47, 48] from $q = 0$ to $q = q_0$ is approximately given by (see also SM[27]):

$$q(t) \simeq \frac{q_0}{2}\left[1 + \tanh\left(\frac{t}{\tau_{tr}}\right)\right], \text{with} \quad \tau_{tr} \simeq \frac{0.34}{\sqrt{\lambda}\sigma}. \tag{8}$$

Finally, to relate this solution to the experimental data shown in Fig. 1, one has to connect $q$ with the shape parameter $\Lambda$. Our microscopy imaging projection, by contrast to our theory, cannot distinguish between prolate $(q > 0)$ and oblate $(q < 0)$ shapes. For a tissue shape similar to the surface of revolution described by Eq. (5), the shape parameter is $\Lambda = 27\left(q^2 - q^3\right)/8 - 405q^4/32 + \cdots$, i.e. $\Lambda \propto q^2$ for $|q| < 1$.

The parameter that characterizes the transition is the time scale, $\tau_{tr}$ which is inversely proportional to the square root of the barrier height $\lambda$, and the noise strength $\sigma$. We do not know how *Heptanol* and electric fields affect the barrier height. However, the noise that drives the transition emerges from the stochastic nature of the $Ca^{2+}$ activity that induces the contractile actomyosin forces within the tissue. These fluctuations are weakened by *Heptanol* and strengthened by the electric field[12]. Assuming a weak dependence of $\lambda$ on the controls implies that $\tau_{tr}$ increases under *Heptanol* while decreases under an electric field. This prediction is consistent with our experimental findings, as demonstrated in Figs. 1d and 1e, and provides support for the first-order nature of the morphological transition.

The main prediction of our model is that the local $Ca^{2+}$ activity is negatively correlated with the scalar curvature. From the coupling action (2), it follows that fluctuations which simultaneously lower the scalar curvature and enhance the $Ca^{2+}$ activity at the same point, reduce the action and are, therefore, more probable.

Our projected microscopy images do not allow a direct measurement of the local curvature of the 3D morphology of the tissue. However, the curvature-$Ca^{2+}$ correlations can be deduced indirectly by measuring the dimensionless second moment of the $Ca^{2+}$ field, $\phi(\boldsymbol{r})$, defined by:

$$M_2 = \frac{\oiint d^2s\left(\boldsymbol{r}-\boldsymbol{r}_c\right)^2\left[\phi(\boldsymbol{r}) - \bar{\phi}\right]}{\oiint d^2s\left(\boldsymbol{r}-\boldsymbol{r}_c\right)^2\bar{\phi}} \ . \tag{9}$$

Here the integral is over the tissue's surface, $\boldsymbol{r}_c$ is its center of mass,

$$\bar{\phi} = \frac{1}{A}\oiint d^2s\phi(\boldsymbol{r}) \tag{10}$$

is the spatial average of $\phi(\boldsymbol{r})$, and $A$ is the surface area. Approximating the tissue's shape by a prolate spheroid, $M_2 > 0$ if the $Ca^{2+}$ activity is statistically enhanced near regions of large curvature, i.e., near the spheroid's poles (Fig. 3a), while $M_2 < 0$ for enhancement near the ellipsoid's equator where the curvature is small (Fig. 3b). Thus, by computing $M_2$ one can deduce the sign of the curvature-$Ca^{2+}$ correlations. An approximate formula for $M_2$ which utilizes the information in the projected image is derived in the SM[27].

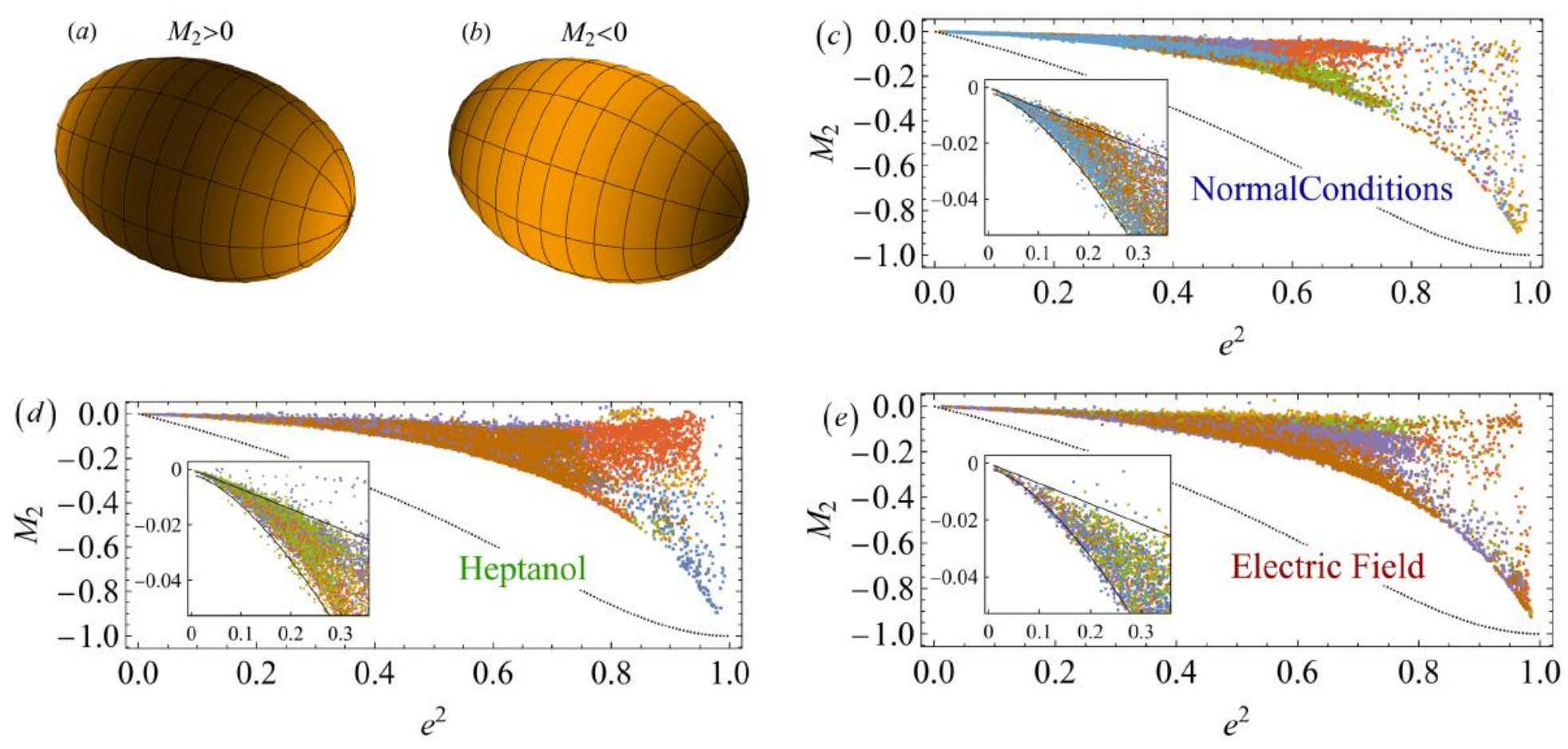

**FIG. 3: The negative associations between the $Ca^{2+}$ field and the tissue's scalar curvature.** (a-b) Illustrations of the average $Ca^{2+}$ activity pattern that leads to positive (a) and negative (b) values of the normalized second moment $M_2$ of the $Ca^{2+}$ field. A positive $M_2$ characterizes an enhanced activity near the poles of a prolate spheroid, while a negative $M_2$ is associated with an enhanced activity near its equator. (c) A scatter plot of $M_2$ as a function of the square spheroid eccentricity $e^2$. Each color represents a different tissue sample (n=7). The inset shows a magnified view near the origin (close to a spheroidal shape). (d-e) The same for tissue samples subjected to *Heptanol* (d; n=6; 300 μl/l in HM) and an electric field (e; n=6; 30V/4mm in HM). The solid lines in the insets are a guide for the eye. The curves in all figures show the upper and lower bounds of $M_2$ (see SM[27]). The upper curves in the figures are the straight lines $M_2 = -e^2/14$, while the lower curves are the inverted parabolas $M_2 = -0.004 - e^4/1.6$.

For each frame along our microscopy imaging trace, we identify the optimal ellipsoid that fits the tissue's projected contour, extract its eccentricity $e$ and compute the corresponding value of $M_2$. The scatter plots of this data is presented in Fig. 3, for all the tissue samples in our experiments under the various controls. In all cases, as the eccentricity increases (i.e., the spheroid becomes more prolate), $M_2$ becomes more negative. Since a more prolate spheroid implies that most of the (positive) scalar curvature is located at the tips of the spheroid, the behavior in Figs. 3d-f indicates that, indeed, the $Ca^{2+}$ activity is negatively correlated with the curvature.

The consistency of the above theoretical description of *Hydra* morphogenesis as a phase transition, driven by the coupling between the $Ca^{2+}$ field and the curvature, leads to two additional insights. The first follows from our hypothesis that the transition of the tissue's morphology into a cylindrical shape emerges from a modulation in the balance between the active actomyosin forces, driven by the $Ca^{2+}$ activity, and the internal cavity fluid pressure. Under this assumption, weakening of the active contractile forces by *Heptanol* should lead to the collapse of the cylindrical-shaped tissue back into a spheroidal shape, reflecting the isotropic nature of the pressure force. Fig. 4a shows that this is indeed the case (see also Movie 3[27]). This result implies that maintaining an elongated cylindrical shape requires a significant distribution of the active internal contractile forces.

For the second insight, recall that *Heptanol* and an external electric field, applied separately, halt regeneration differently: The first weakens the $Ca^{2+}$ activity while the latter strengthens it[12]. If the $Ca^{2+}$ fluctuations are the primary agent driving the morphological transition, then applying both controls simultaneously for a tissue halted by *Heptanol*, such that the $Ca^{2+}$ activity is resumed to be in a proper range, should reinitiate morphogenesis. The experimental data shown in Figs. 4b-c demonstrates the validity of this prediction (see also Movie 4[27]). Note that the $Ca^{2+}$ activity (measured here by its spatial mean) is synchronized with the externally applied modulated voltage, demonstrating that at this range of parameters, the voltage indeed enhances the $Ca^{2+}$ activity, which is otherwise suppressed under *Heptanol*.

In summary, concentrating on the morphological transition in whole-body *Hydra* regeneration, we develop a physical model for morphogenesis. The negative correlations between the $Ca^{2+}$ spatial fluctuations and the local tissue curvature suggest a simple picture of this transition: Enhanced contractile forces at regions of low curvature lead to elongation of the spheroidal shape of the tissue by the internal cavity pressure force. The transition is further driven along the body-axis until the tissue's shape is transformed from a spheroid into a cylinder[7, 20]. This simple picture reminds the forms and shapes created in children's party balloons.

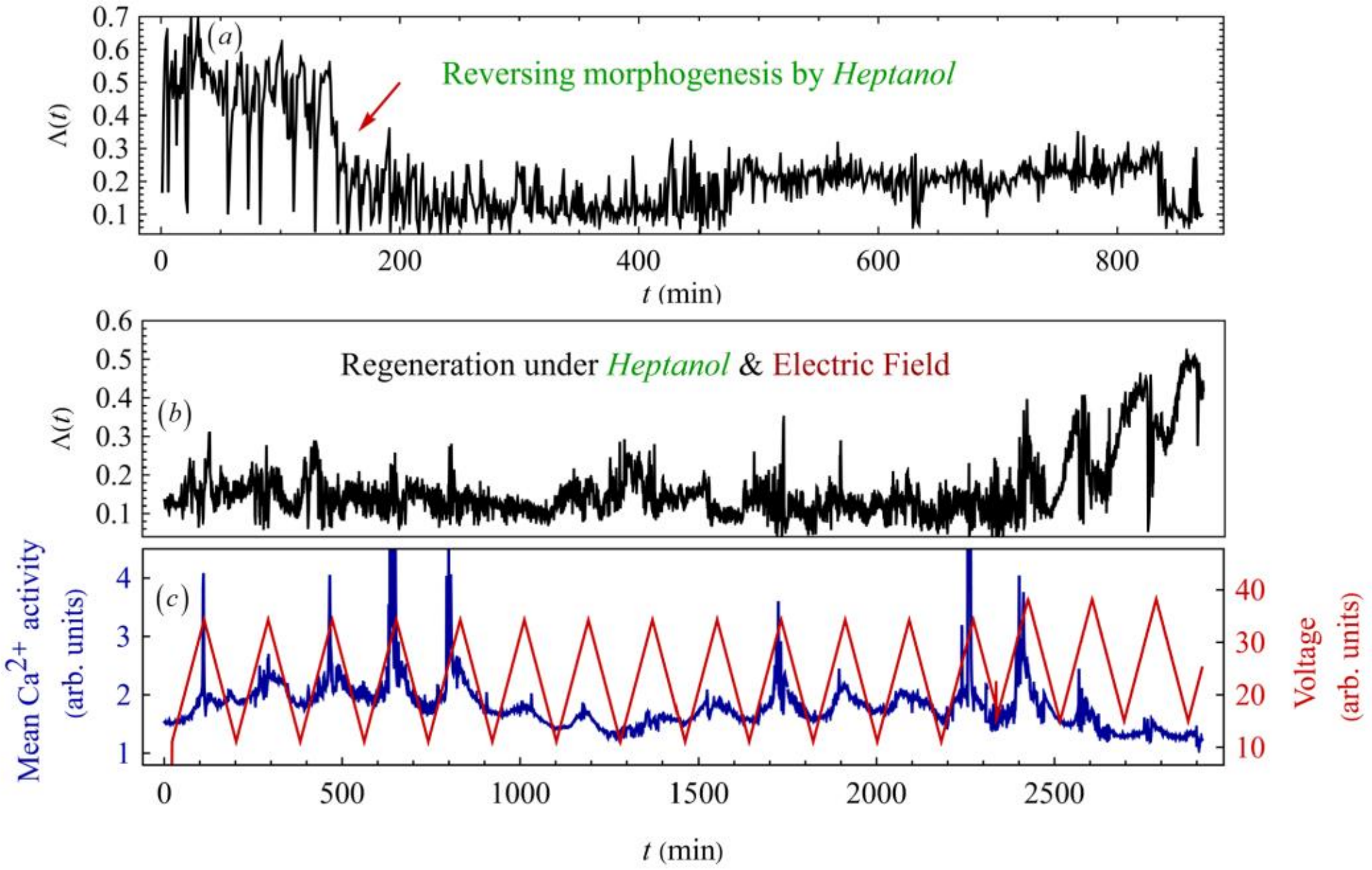

**FIG. 4: Reversing morphogenesis by *Heptanol* and re-inducing it by an electric field.** (a) An example of a time trace of the shape parameter $\Lambda$ of a regenerated mature *Hydra* subjected to *Heptanol* (700 μl/l in HM). The trace clearly shows the folding of the *Hydra*'s body-form from a cylindrical shape (high $q^2$ ) into a spheroidal one (low $q^2$ ). This result was reproducible in 3 separate experiments where a complete trace, similar to the one shown here, were extracted from each of them. It demonstrates that the application of *Heptanol*, blocking gap-junctions and weakening the $Ca^{2+}$ activity, may stabilize the minimum of $V(q)$ at $\Lambda = q = 0$ which is associated with the spherical shape. (b) An example of a time trace of the shape parameter $q^2$ of a spheroidal-shape *Hydra*'s tissue, exhibiting halted regeneration for a long duration (around 60 hr prior to the shown trace) under *Heptanol* (300 μl/l in HM), following the application of an external electric field (at t=0; red curve) in the presence of the *Heptanol*. At these parameters of the controls, each one of them separately halts regeneration and stabilizes a spheroidal-shaped tissue. The simultaneous application of both controls reinitiates the regeneration process and leads to a morphological transition into a cylindrical shape (manifested by a significant increase in $\Lambda$ ). (c) The time traces of the applied voltage (red) and the mean $Ca^{2+}$ activity (blue) corresponding to the time trace of (b). These traces demonstrate the synchronization of the $Ca^{2+}$ activity with the voltage, which is also observed in the evolution of the hydra shape parameter $\Lambda(t)$ . Similar results were observed in 2 separate experiments.

Under *Heptanol*, the $Ca^{2+}$ activity is too weak to provide the necessary contractile forces, so morphogenesis is halted in the spheroidal state. However, the reason for samples subjected to a strong enough electric field to be halted as well is puzzling. Our model provides an explanation for this phenomenon. As shown by us before, the electric field increases the $Ca^{2+}$ activity and its correlation length making its distribution more uniform over the tissue[12]. We have shown that the electric field also leads to enhanced temporal frequencies of the $Ca^{2+}$ fluctuations above the mechanical relaxation time of the tissue[12]. Thus, the slow component of $Ca^{2+}$ activity becomes more uniform over the tissue. A uniform $Ca^{2+}$ distribution cannot provide the localized forces required for a morphological change. From a mathematical viewpoint, the electric field reduces the coupling strength between the $Ca^{2+}$ field and the local curvature, since for a uniform $\phi$ the action (2) becomes a topological invariant. Another reason is that, as in any biological system, the coupling action as a function of $\phi$ features a saturation (see Eq. (D.12)[27]). Thus, increasing the average $Ca^{2+}$ activity drives the system into a regime where the force gradients become smaller and cannot provide the localized contractile forces required for the morphological transition.

It remains for future work to understand the detailed mechanisms leading to the feedback between curvature and the $Ca^{2+}$ activity as well as the details of how this activity is reflected in the distribution of the internal active actomyosin forces. Finally, the body-axis alignment, dictated by the longitudinal supracellular actin fibers, and the axis polarity which determines the position of the head in the regenerating tissue segment, are both strongly inherited from the parent *Hydra*[7, 20, 49-53]. An important open issue is the effects of these symmetry-breaking fields on morphogenesis.

**Acknowledgments**

We thank Eldad Bettelheim, Benny Davidovitch, Omri Gat, Dror Orgad, Naama Brenner, Kinneret Keren, Yariv kafri, Guy Bunin and Shimon Marom, for discussions and comments on the manuscript. EB thanks the lab members: Liora Garion and Yonit Maroudas-Sacks for their technical help. Special thanks to Gdalyahu Ben-Yoseph for superb technical help in designing and constructing the experimental setup, and to Anatoly Meller for constructing the electrical control system. This work was supported by a grant (EB) from the Israel Science Foundation (Grant no. 1638/21)).

**References**

1 J. B. A. Green and J. Sharpe, Development **142**, 1203 (2015).
2 P. A. Lawrence, Nature Cell Biology **3**, E151 (2001).
3 E. Braun and K. Keren, BioEssays **40**, 1700204 (2018).
4 C. J. Chan, C.-P. Heisenberg, and T. Hiiragi, Current Biology **27**, R1024 (2017).
5 T. Lecuit, P. F. Lenne, and E. Munro, Annu. Rev. Cell Dev. Biol. **27**, 157 (2011).
6 T. Lecuit and L. Mahadevan, Development **144**, 4197 (2017).
7 A. Livshits, L. Shani-Zerbib, Y. Maroudas-Sacks, et al., Cell Reports **18**, 1410 (2017).
8 C. J. Miller and L. A. Davidson, Nature Rev. Genetics **14**, 733 (2013).
9 T. Mammoto, A. Mammoto, and D. E. Ingber, Annual Review of Cell and Developmental Biology **29**, 27 (2013).
10 E. Hannezo and C.-P. Heisenberg, Cell **178** (2019).
11 S. P. Banavar, E. K. Carn, P. Rowghanian, et al., Scientific Reports **11**, 8591 (2021).
12 O. Agam and E. Braun, BioRxiv: https://doi.org/10.1101/2021.11.01.466811; arxiv: https://doi.org/10.48550/arXiv.2303.02671 (2023).
13 E. Braun and H. Ori, Biophysical Journal **117**, 1514 (2019).
14 M. Levin, Cell **184**, 1971 (2021).
15 M. Levin and C. J. Martyniuk, Biosystems **164**, 76 (2018).
16 B. B. Silver and C. M. Nelson, Frontiers in Cell and Developmental Biology **6** (2018).
17 B. Galliot, Int. J. Dev. Biol. **56**, 411 (2012).
18 B. Galliot, *Regeneration in Hydra; Encyclopedia of Life Sciences* (John Wiley & Sons Ltd, Chicester, 2013).
19 H. R. Bode, Cold Spring Harbor Perspectives in Biology **a000463**, 1 (2009).
20 Y. Maroudas-Sacks, L. Garion, L. Shani-Zerbib, et al., Nature Physics **17**, 251 (2021).
21 P. M. Bode and H. R. Bode, Developmental biology **106**, 315 (1984).
22 C. Futterer, C. Colombo, F. Julicher, et al., EPL **64**, 137 (2003).
23 A. Gierer, S. Berking, H. Bode, et al., Nature/New Biology, 98 (1972).
24 M. Kucken, J. Soriano, P. A. Pullarkat, et al., Biophys J **95**, 978 (2008).
25 H. Meinhardt, Int. J. Dev. Biol. **56**, 447 (2012).
26 H. D. Park, A. B. Ortmeyer, and D. P. Blankenbaker, (1970).
27 SM.
28 R. Aufschnaiter, R. Wedlich-SÃ, X. Zhang, et al., Biology Open **6**, 1137 (2017).
29 N. P. Mitchell, D. J. Cislo, S. Shankar, et al., eLife **11**, e77355 (2022).
30 M. F. Lefebvre, N. H. Claussen, N. P. Mitchell, et al., eLife **12**, e78787 (2023).
31 S. J. Streichan, M. F. Lefebvre, N. Noll, et al., eLife **7**, e27454 (2018).
32 J. Szymanski and R. Yuste, Current Biology **29**, 1807 (2019).
33 C. Duclut, J. Prost, and F. Julicher, Proceedings of the National Academy of Sciences **118**, e2021972118 (2021).
34 M. Popović, J. Prost, and F. Jülicher, arXiv:2211.06909 (2022).
35 J. Soriano, S. Rudiger, P. Pullarkat, et al., Biophys J **96**, 1649 (2009).
36 A. Hernandez, M. F. Staddon, M. Moshe, et al., arXiv:2303.06224v1 (2023).
37 N. Murisic, V. Hakim, IoannisÂ G. Kevrekidis, et al., Biophysical Journal **109**, 154 (2015).
38 D. Khoromskaia and G. Salbreux, eLife **12**, e75878 (2023).
39 R. Jackiw, edited by A. H. Bristol, 1984).
40 C. Teitelboim, edited by A. H. Bristol, 1984).
41 E. Farge, Current Biology **13**, 1365 (2003).
42 C. Guillot and T. Lecuit, Science **340**, 1185 (2013).
43 A. Zakharov and K. Dasbiswas, The European Physical Journal E **44**, 82 (2021).
44 A. Zakharov and K. Dasbiswas, Soft Matter **17**, 4738 (2021).
45 P. Canham, Journal of Theoretical Biology **26**, 61 (1970).
46 W. Helfrich, Zeitschrift für Naturforschung, **28c**, 693 (1973).
47 U. Weiss, Phys. Rev. A **25**, 2444(R) (1982).
48 K. L. C. Hunt and J. Ross, J. Chem. Phys **75**, 976 (1981).
49 A. Livshits, L. Garion, Y. Maroudas-Sacks, et al., Scientific Reports **12**, 13368 (2022).
50 L. Shani-Zerbib, L. Garion, Y. Maroudas-Sacks, et al., Genes **13**, 360 (2022).

[51] H. R. Bode, Int. J. Dev. Biol. **56**, 473 (2012).
[52] R. Wang, R. E. Steele, and E.-M. S. Collins, Developmental Biology **467**, 88 (2020).
[53] H. K. MacWilliams, Developmental biology **96**, 239 (1983).

# Supplementary Material for

# *Hydra* morphogenesis as phase-transition dynamics

Oded Agam[1] and Erez Braun[2]

[1]*The Racah Institute of Physics, Edmond J. Safra Campus, The Hebrew University of Jerusalem, Jerusalem 9190401, Israel.*
[2] *Department of Physics and Network Biology Research Laboratories, Technion-Israel Institute of Technology, Haifa 32000, Israel.*

## Content

*A. Movies*

**https://www.dropbox.com/sh/5xxd45a599u8lol/AAAX0AizdvUIoHTf-5NLcSjCa?dl=0**

**Movie 1: *Hydra* regeneration from a small tissue segment.** Bright-field (BF) images of a tissue segment, after its folding into a spheroid, undergoing a complete regeneration process into a mature *Hydra*.

**Movie 2: The morphological transition.** Pairs of bright-field (BF) and fluorescence images of the same tissue spheroid as in Movie 1, around the morphological transition from a spheroidal shape into a cylindrical one. Note the sharpness of the transition.

**Movie 3: Reversing morphogenesis by *Heptanol*.** Bright-field (BF) images of the tissue sample shown in Fig.4a in the main text. It demonstrates the folding of a mature *Hydra* into a spheroidal shape under the application of *Heptanol*. See the caption of Fig.4 and Methods for details.

**Movie 4: Re-initiation of regeneration under the simultaneous application of *Heptanol* and an electric field.** Bright field (BF) images of a tissue sample shown in Figs.4b-c in the main text. The tissue is maintained in the spheroidal shape for a long duration (~60 hrs) under *Heptanol* that halts regeneration (only a section of this duration is shown). An electric field is then applied to the sample, leading to re-initiation of the regeneration process; first, a transition from a spheroidal to a cylindrical shape and eventually to the emergence of tentacles and a mature *Hydra*. See the caption of Fig.4 and Experimental Methods for details.

*B. Experimental Methods*

Experiments are carried out with a transgenic strain of *Hydra Vulgaris* (*AEP*) carrying a GCaMP6s probe for $Ca^{2+}$ (see Ref. [1, 2] for details of the strain). Animals are cultivated in a *Hydra* culture medium (HM; 1mM $NaHCO_3$, 1mM $CaCl_2$, 0.1mM $MgCl_2$, 0.1mM KCl, 1mM Tris-HCl pH 7.7) at 18°C. The animals are fed every other day with live *Artemia nauplii* and washed after ~4 hours. Experiments are initiated ~24 hours after feeding. Tissue segments are excised from the middle of a mature *Hydra*. Fragments are incubated in a dish for ~3 hrs to allow their folding into spheroids prior to transferring them into the experimental sample holder. For the experiments in *Heptanol*: the flowing medium is replaced with HM containing 300 μl/l of *Heptanol* (1- Heptanol, 99%, Alfa Aesar) after good mixing. For many tissue samples, this concentration of *Heptanol* halts regeneration. Approximately 20-30% of the samples in each experiment escape the *Heptanol* halting and regenerate into a fully developed mature *Hydra*. These are the tissue samples analyzed in this Letter. For the reversal of regenerated *Hydra*, the level of *Heptanol* is increased gradually up to 700 μl/l and then maintained at this level for the duration of the experiment. Tissue samples regenerated under an external electric field are observed in the same HM medium under a constant 30V AC field at 2 kHz (4 mm distance between the electrodes). This applied electric field is below the critical field halting regeneration, so the tissue samples, although affected by the field as observed in the $Ca^{2+}$ signal, can still regenerate into a fully developed mature *Hydra*. For tissue samples regenerated under both controls, the samples are observed under 300 μl/l *Heptanol* for approximately 60 hrs, and then an AC voltage at 2 kHz, alternating in the form of a triangular shape at a periodicity of 3 hr is applied, first in the range of 10-40 V and then increased to the range 15-45V. Full regeneration emerges after the application of the increased voltage.

The experimental setup is similar to the one described in Ref[1]. In all the experiments, spheroid tissues are placed within wells of ~1.3 mm diameter made in a strip of 2% agarose gel (Sigma) to keep the regenerating *Hydra* in place during time-lapse imaging. The tissue spheroids, typically of a few hundred microns in size, are free to move within the wells. The agarose strip containing 15 wells, is fixed on a transparent plexiglass bar of 1 mm height, anchored to a homemade sample holder. Each well has two platinum mesh electrodes (Platinum gauze 52 mesh, 0.1 mm dia. Wire; Alfa Aesar, Lancashire UK) fixed at its two sides at a distance of 4 mm between them, on two ceramic-filled polyether ether ketone (CMF Peek) holders. A channel on each side separates the sample wells from the electrodes allowing for medium flow. Each electrode pair covers the entire length of the well and its height, ensuring full coverage of the tissue sample. A peristaltic pump (IPC, Ismatec, Futtererstr, Germany) flows the medium (either HM or HM+*Heptanol*) continuously from an external reservoir (replaced at least once every 24 hrs) at a rate of 170 ml/hr into each of the channels between the electrodes and the samples. The medium covers the entire preparation, and the volume in the bath is kept fixed throughout the experiments by pumping the medium out from 4 holes determining the fluid's height. The continuous medium flow ensures stable environmental conditions, and the fixed volume of medium in the bath ensures constant conductivity between the electrodes (measured by the stable current between the electrodes when voltage is applied). All the experiments are done at room temperature.

Time-lapse bright-field and fluorescence images are taken by a Zeiss Axio-observer microscope (Zeiss, Oberkochen Germany) with a 5× air objective (NA=0.25) and a 1.6× optovar and acquired by a CCD camera (Zyla 5.5 sCMOS, Andor, Belfast, Northern Ireland). The sample holder is placed on a movable stage (Marzhauser, Germany), and the entire microscopy system is operated by Micromanager, recording images at 1 min intervals. The fluorescence recording at 1 min resolution is chosen on the one hand to allow long experiments while preventing tissue damage throughout the experiments and, on the other hand, to enable recordings from multiple tissue samples.

*C. Time traces of individual samples.*

Time traces of the shape parameter, $\Lambda(t)$, and the projected area (normalized by its time average) of individual tissue samples used to calculate Fig. 1 are presented in Figs. S1, S2, and S3; under normal conditions, under an electric field, and under *Heptanol*, respectively. The black curves in these figures represent the shape parameter, while the colored curves are traces of the projected area. In all three cases, the area typically grows after the morphological transition. This area growth is not associated with cell division which occurs on much longer time scales – of the order of a day or few days[3-5].

In the normal case, the tissue's projected area exhibits sawtooth-like dynamics. This pattern is believed to be driven by the osmotic pressure gradients, leading to a relatively smooth inflation of the tissue due to water influx, followed by a sudden collapse of the tissue due to local rupture[6-8]. This feature is unlikely to be directly related to the morphological transition because it is suppressed in samples subjected to an electric field that nevertheless regenerate. Moreover, the sawtooth pattern is typically suppressed before the transition, and as was shown eriler  this suppression does not mark any significant event in the body-axis formation along regeneration[9].

*D. The model and its extensions.*

The field theoretic model presented in this work contains three terms. The first is the action of the $Ca^{2+}$ field given by Eq. (1). This action is obtained from an equilibrium-like solution of the Fokker-Planck equations derived from the dynamical stochastic equation of a self-excitatory field, and its validity has been verified in our previous work[1]. The other two terms are the coupling (2), and the shape (3) actions. The purpose of this section is to provide support for this model. To this end, we first derive the coupling action from microscopic considerations and discuss additional coupling terms that are neglected in our model. Then, we comment about: (a) the shape action (3), (b) the justification for considering the system to be two-dimensional, and (c) the feedback of the local curvature on the calcium activity.

In understanding the coupling action (2), one should first notice that the $Ca^{2+}$ field is directly coupled to the local tissue structure via the contractile actomyosin fibers, which form a crisscross mesh in the two epithelial layers of the tissue, as shown in Fig. S4. The $Ca^{2+}$ activity enables the operation of the myosin motors on the actin filaments generating active contractile forces in the tissue. The supracellular actin fibers are located on both sides of the extracellular matrix separating the ectoderm (outer) and endoderm (inner) bilayer epithelium. Equal contraction of the actin fibers on both sides of the extracellular matrix decreases the intrinsic (scalar) curvature, $\mathcal{R}$, and keeps the extrinsic (mean) curvature, $\mathcal{K}$, unchanged as demonstrated in the lower panels of  Fig. S4. Thus, the primary coupling term of the $Ca^{2+}$ activity to the tissue's geometrical properties is via the scalar curvature.

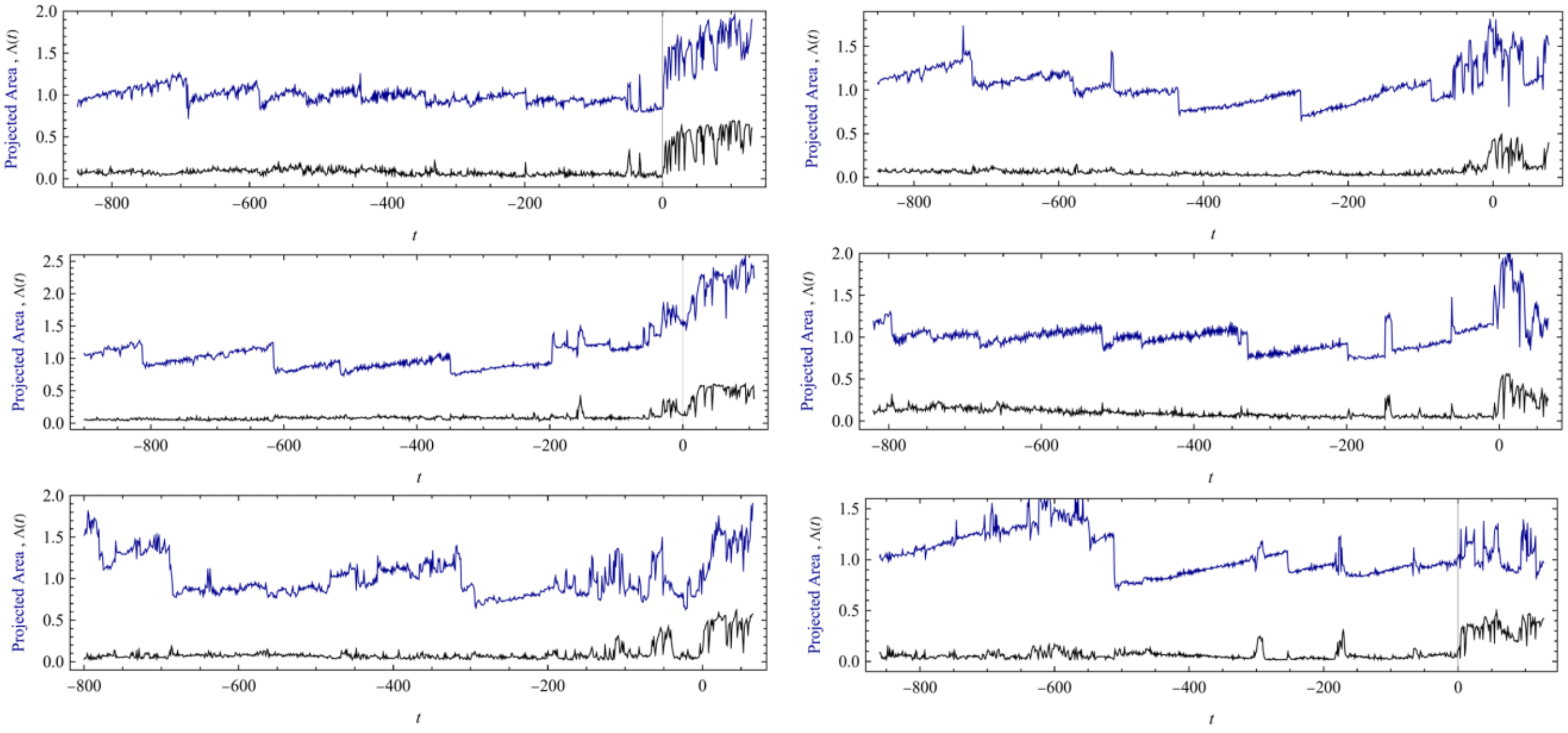

**FIG. S1:** Time traces of the shape parameter (black) and the projected area normalized by its time average (blue), of individual sample under normal conditions. The shape parameter is defined by $\Lambda = 1 - 4\pi A / L^2$ where $A$ is the area of the projected tissue, while $L$ is its perimeter. The time axis was shifted such that $t = 0$ marks the onset of the morphological transition of the tissue from a spheroidal shape into a cylindrical one.

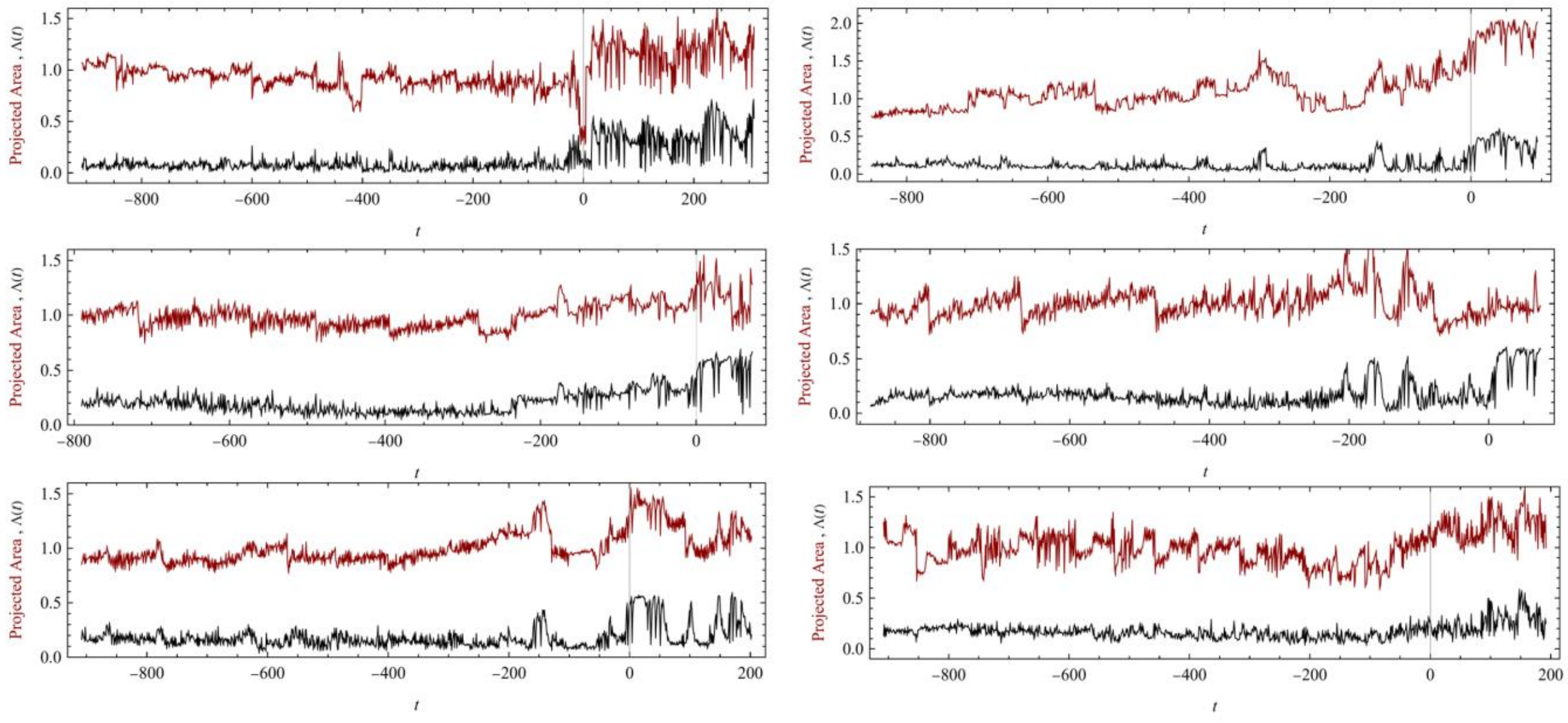

**FIG. S2:** The same as Fig. S1 for tissue samples under an electric field.

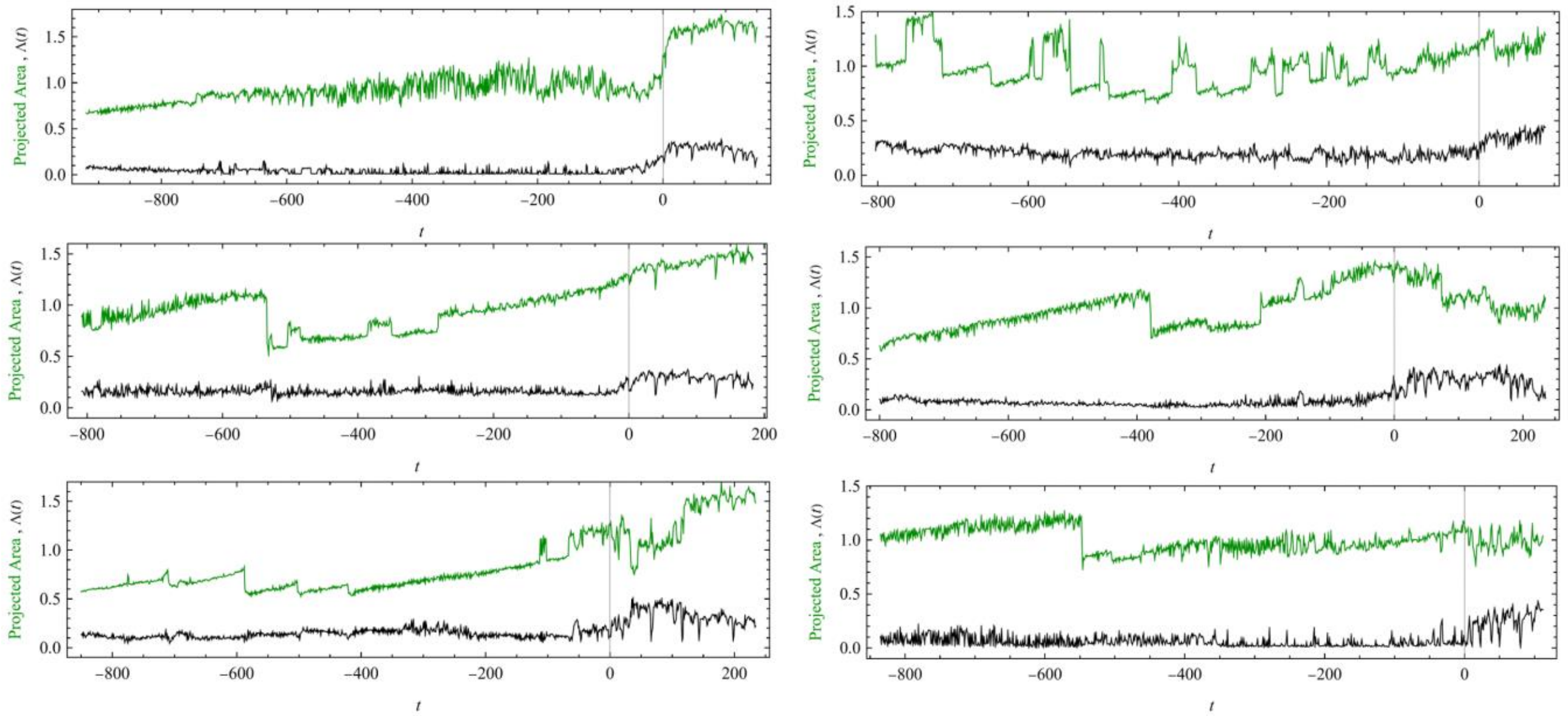

**FIG. S3:** The same as Fig. S1 for tissue samples under *Heptanol*.

We turn now to derive this result and construct additional coupling terms of the $Ca^{2+}$ activity field, $\phi$, to the nematic order of the actin fibers and the polarity of the tissue. The supracellular actin fibers are arranged in a nematic order within each tissue layer, and this nematic order is relatively smooth (outside of a few local defects) and highly stable[10]. Also, the tissue segment carries an inherited polarity from the parent animal[11], which eventually dictates the side of the head formation in the regenerating *Hydra*, i.e., the polarity of the body-axis of the regenerated animal. We argue below that these additional terms play a minor role in the morphological transition.

Let $\hat{\boldsymbol{t}}$ denotes a tangent unit vector to the *Hydra*'s tissue surface, oriented along the fibers of the outer epithelial layer (ectoderm). We define the tensor field,

$$\tilde{Q}^{\text{out}}_{\mu\nu} = \left\langle \hat{t}_\mu \hat{t}_\nu \right\rangle , \qquad \text{(D.1)}$$

where the averaging is over a small region (compared to the tissue size) that contains many fibers. Since $\text{Tr}\tilde{Q}^{\text{out}} = 1$ it is convenient to define the traceless tensor $Q^{\text{out}} = \tilde{Q}^{\text{out}} - I/2$, where $I$ is the identity matrix. For instance, in a flat space, with $\hat{\boldsymbol{t}} = (\cos\theta, \sin\theta)$, we have

$$Q^{\text{out}}_{\mu\nu} = \begin{pmatrix} \left\langle \cos^2\theta \right\rangle - \frac{1}{2} & \left\langle \cos\theta \sin\theta \right\rangle \\ \left\langle \cos\theta \sin\theta \right\rangle & \left\langle \sin^2\theta \right\rangle - \frac{1}{2} \end{pmatrix} = \frac{1}{2}\begin{pmatrix} \left\langle \cos 2\theta \right\rangle & \left\langle \sin 2\theta \right\rangle \\ \left\langle \sin 2\theta \right\rangle & -\left\langle \cos 2\theta \right\rangle \end{pmatrix}, \qquad \text{(D.2)}$$

and the nematic order parameter is defined as:

$$q = \sqrt{\left\langle \cos^2 2\theta \right\rangle + \left\langle \sin^2 2\theta \right\rangle} , \qquad \text{(D.3)}$$

The nematic director is determined by the vector $\hat{\boldsymbol{e}}$, which satisfies the relation:

$$Q^{\text{out}} = q\left[ \hat{\boldsymbol{e}} \otimes \hat{\boldsymbol{e}} - \frac{1}{2} I \right] \qquad \text{(D.4)}$$

Namely if $\varphi = \cos^{-1}\left( \left\langle \cos 2\theta \right\rangle / q \right)$, then $\hat{\boldsymbol{e}} = (\cos\varphi, \sin\varphi)$. A similar expression characterizes the nematic order parameter of the inner fiber mesh (the endoderm), which we denote by $Q^{\text{in}}$. For simplicity, we assume that the nematic order parameters, $q$, of the inner and outer layers are the same.

From the scalar field $\phi$, the tensors $\tilde{Q}^{\text{out}}_{\mu\nu}$ and $\tilde{Q}^{\text{in}}_{\mu\nu}$, the intrinsic curvature tensor, $\mathcal{R}^{\mu\nu}$, and the extrinsic curvature tensor, $\mathcal{K}^{\mu\nu}$, one can construct several scalars describing the coupling of the $Ca^{2+}$ field to the tissue's geometrical properties. The most obvious ones are the direct coupling of $\phi$ to the scalar and mean curvatures:

$$\eta^{\mathcal{R}}_{\text{in}} \phi \frac{1}{2} g_{\mu\nu} \mathcal{R}^{\nu\mu} + \eta^{\mathcal{R}}_{\text{out}} \phi \frac{1}{2} g_{\mu\nu} \mathcal{R}^{\nu\mu} = \eta \phi \mathcal{R} , \qquad \text{(D.5)}$$

and

$$-\eta^{\mathcal{K}}_{\text{in}} \phi \frac{1}{2} g_{\mu\nu} \mathcal{K}^{\nu\mu} + \eta^{\mathcal{K}}_{\text{out}} \phi \frac{1}{2} g_{\mu\nu} \mathcal{K}^{\nu\mu} = \frac{\Delta\eta}{2} \phi \mathcal{K} , \qquad \text{(D.6)}$$

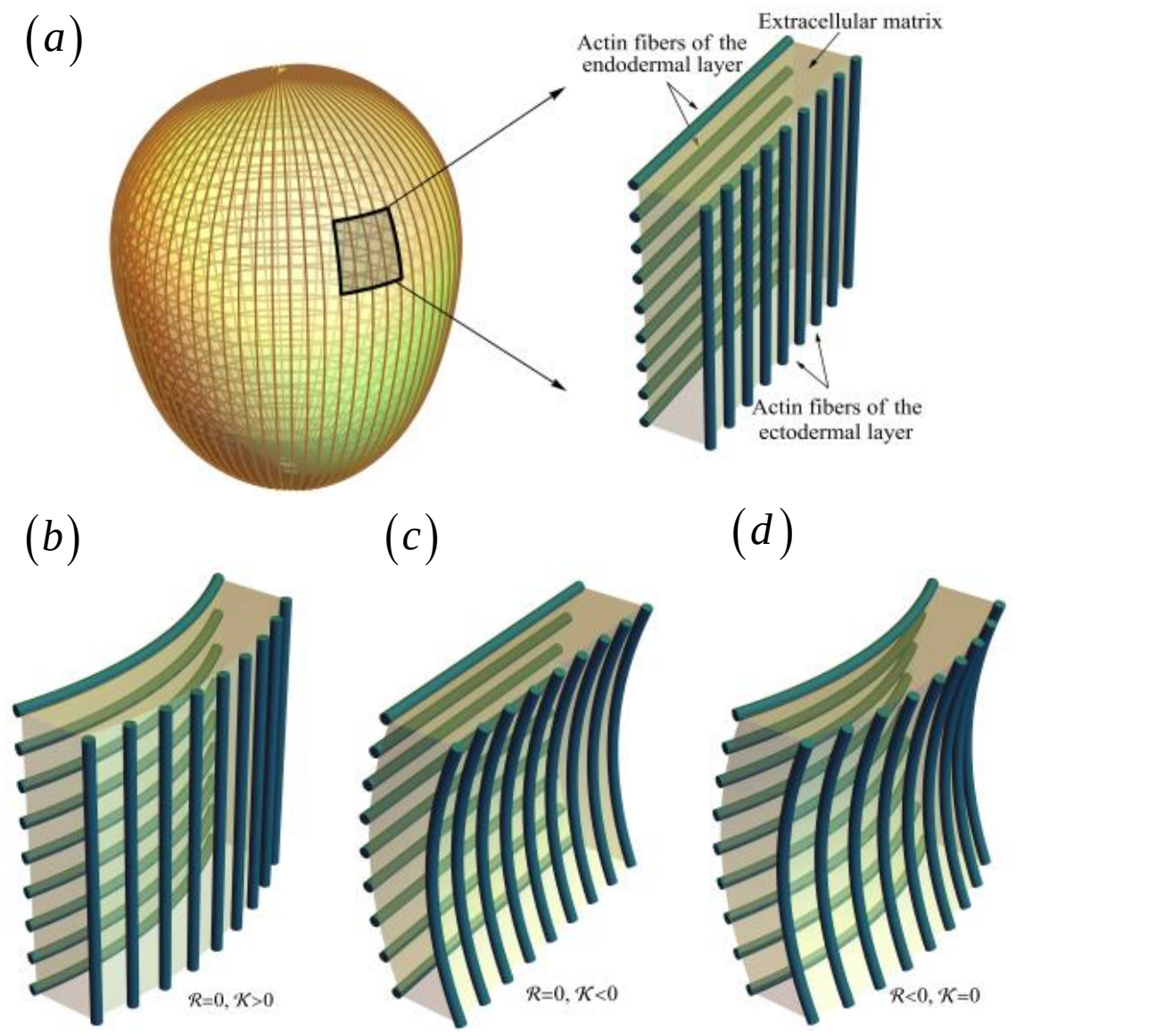

**FIG. S4:** (a) The actin fibers mesh of the *Hydra* tissue. (b) and (c) the effect of contraction of fibers either in the endodermal or ectodermal tissue layers that changes the mean curvature and leaves the scalar curvature intact. (d) Contractions of actin fibers in both tissue layers reduce the scalar curvature but do not change the mean curvature.

where $g_{\mu\nu}$ is the metric tensor (see definition in section F below), and $\eta^{\mathcal{K}/\mathcal{R}}_{\text{in/out}}$ are the coupling constants of the mean and scalar curvatures to the inner and outer layers. These coupling constants add up for the scalar curvature, $\eta = \left(\eta^{\mathcal{R}}_{\text{out}} + \eta^{\mathcal{R}}_{\text{in}}\right)/2$, while they are subtracted for the mean curvature, $\Delta\eta = \eta^{\mathcal{K}}_{\text{out}} - \eta^{\mathcal{K}}_{\text{in}}$. Eq. (D.5) leads to Eq. 2. The coupling of the $Ca^{2+}$ field to the mean curvature $\phi\mathcal{K}$ vanishes when $\eta^{\mathcal{K}}_{\text{out}} = \eta^{\mathcal{K}}_{\text{in}}$. Moreover, it is expected to be smaller than the coupling to the intrinsic curvature since the mean curvature may not change significantly during the morphological phase transition, while the scalar curvature must change. Hence, viewing the $Ca^{2+}$ activity as a significant factor in driving the phase transition, one expects that this field couples primarily to the scalar curvature.

Consider now the coupling of the scalar curvature tensor to the nematic order parameter. Let $\hat{\boldsymbol{e}}$ denotes a three dimensional vector associated with the nematic director (of either the inner or the outer layers), which is tangent to the *Hydra*'s tissue surface, then

$$\begin{aligned} \phi Q_{\mu\nu}\mathcal{R}^{\mu\nu} &= \phi q\left(e_\mu e_\nu - \frac{1}{2}g_{\mu\nu}\right)\mathcal{R}^{\mu\nu} \\ &= \phi q\left(e_\mu e_\nu - \frac{1}{2}g_{\mu\nu}\right)\frac{\mathcal{R}}{2}g^{\mu\nu} = \phi q\frac{\mathcal{R}}{2}\left(e_\mu e^\mu - \frac{1}{2}\delta^\mu_\mu\right) = 0 \end{aligned} \tag{D.7}$$

where $g_{\mu\nu}$ is the metric tensor. To obtain the second line in the above equation, we used the property $2\mathcal{R}^{\mu\nu} = \mathcal{R}g^{\mu\nu}$ that holds for two-dimensional surfaces. Thus, independent of the values of the coupling constants $\eta^{\mathcal{R}}_{\text{out}}$ and $\eta^{\mathcal{R}}_{\text{in}}$ the scalar curvature tensor does not couple to the nematic order parameter.

On the other hand, assuming $\eta^{\mathcal{K}}_{\text{out}} = \eta^{\mathcal{K}}_{\text{in}} = \eta_n$, and that the actin fibers in the inner tissue layer are locally perpendicular to those of the outer layer, $Q^{\text{in}}_{\mu\nu} = -Q^{\text{out}}_{\mu\nu}$ (as can be verified by rotating $\theta \to \theta + \pi/2$ in (D.2)), we see that the coupling of the nematic order parameter to the extrinsic curvature tensor takes the form:

$$\begin{aligned} S_{\text{nematic},\phi} &= \oiint d^2x\sqrt{g}\phi\left(\eta^{\mathcal{K}}_{\text{out}}Q^{\text{out}}_{\mu\nu} - \eta^{\mathcal{K}}_{\text{in}}Q^{\text{in}}_{\mu\nu}\right)\mathcal{K}^{\mu\nu} \\ &= \eta_n \oiint d^2x\sqrt{g}\phi q\left(2e^{\text{out}}_\mu e^{\text{out}}_\nu - g_{\mu\nu}\right)\mathcal{K}^{\mu\nu} \end{aligned} \tag{D.8}$$

On a sphere with actin fibers (of the outer layer) that follow the longitudinal lines (i.e., $\boldsymbol{e}^{\text{out}}$ parallel to the longitudinal lines), one obtains:

$$\phi\left(\eta_{\text{out}}Q^{\text{out}}_{\mu\nu} - \eta_{\text{in}}Q^{\text{in}}_{\mu\nu}\right)\mathcal{K}^{\mu\nu} \xrightarrow[\text{sphere}]{} 0\,, \tag{D.9}$$

while, for a cylinder, with $\boldsymbol{e}^{\text{out}}$ parallel to its axis,

$$\phi\left(\eta_{\text{out}}Q^{\text{out}}_{\mu\nu} - \eta_{\text{in}}Q^{\text{in}}_{\mu\nu}\right)\mathcal{K}^{\mu\nu} \xrightarrow[\text{cylinder}]{} -\eta_n\phi q\mathcal{K}\,. \tag{D.10}$$

Thus, near a spherical shape, the coupling of the $Ca^{2+}$ field to the nematic order parameter is small. Additional reasons that $S_{\text{nematic},\phi}$ may be small are imperfection in the crisscross mesh of the actin fibers and topological defects that reduce the average nematic order parameter. Yet, the coupling of the calcium fluctuations to the nematic order parameter (D.8), even when small, provides a symmetry-breaking term that determines the direction of the elongation of the tissue during regeneration.

As explained above, the *Hydra*'s tissue also maintains information about the polarity of the body-axis, which is inherited from the parent animal[11]. Namely, tissue cells possess polarity information. Denoting by $\boldsymbol{P}$ the unit vector associated with this polarity, the simplest coupling term of the $Ca^{2+}$ activity and this vector is given by the action:

$$S_{\text{polarity},\phi} = \eta_P \oiint d^2x\sqrt{g}P_\mu\nabla^\mu\phi \tag{D.11}$$

This term generates asymmetry between the *Hydra's* head and foot since it produces a gradient of $\phi$ in that direction. However, it does not drive the transition from a spherical into an elongated shape. Notwithstanding, it is likely to play a role in the later stages of the regeneration process.

To colse this discussion over other possible coupling terms between the calcium activity and the tissue's geometrical structure, notice that a linear coupling to the $Ca^{2+}$ activity is an approximation. More generally, one expects the coupling to saturate to some constant value at high values of $\phi$. A more accurate description of the coupling term, Eq. 2, is obtained when $\phi$ is replaced by some nonlinear function $w(\phi)$, such as,

$$w(\phi) = \frac{\phi}{1+\phi/\phi_{sat}}, \quad \text{(D.12)}$$

where $\phi_{sat}$ is the saturation value of the field.

We conclude this section with a few comments. The first is about the shape action given by Eq. (3). We are interested in a coarse-grained description of the whole-tissue morphogenesis, in contrast to other models of epithelial morphogenesis which describe the system at the cell resolution level (e.g., vertex models[12, 13]). The *Hydra* is not a simple lipid membrane nor an elastic sheet. It is, essentially, a muscle made out of cells that actively change their shape, with a thickness that is rather large compared to the tissue size as demonstrated by the schematic cross-section depicted in Fig. S5.

The time traces of the tissue's projected area, presented in Figs. S1-3, show large changes in the tissue's area, especially near the morphological transition. These changes, which do not involve cell division (that happens on much longer time scales)[5, 14], are associated with the contraction of the cells along their long sides and widening in the perpendicular directions. This property implies that stretching and compression are a soft modes of the tissue. In other words, configurations of the tissue associated with large deviations from isometry (i.e., preserving the length of a line on the surface connecting any two points) are not effectively suppressed. Shearing also represents a deviation from isometry; hence the only remaining modes of the tissue are bending modes associated with the scalar and the mean curvature. The scalar curvature integrated over a closed two-dimensional surface is a topological invariant; hence the only contribution to the shape action comes from the mean curvature (neglecting terms that are higher order in the scalar curvature). Taking also into account the natural (spontaneous) curvature of the tissue, $\mathcal{K}_0$, leads to the shape action (3).

The tissue's cross-section shown in Fig. S5 raises a question concerning the validity of our two-dimensional modeling of the system. However, one should notice that the coupling between the $Ca^{2+}$ and the tissue takes place mainly in the vicinity of the thin extracellular matrix separating the two epithelial layers, ectoderm and endoderm, where the supracellular actin fibers reside. Thus, for a description of the morphological transition, one can assume the system to be approximately two-dimensional.

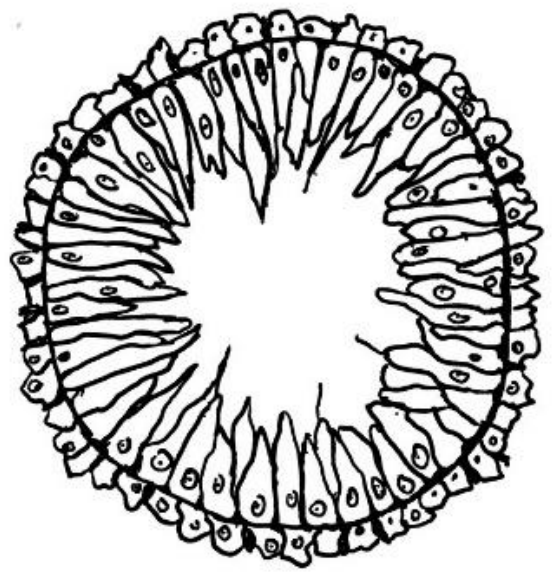

**FIG. S5:** A caricature of *Hydra*'s cross-section. The tissue is composed of two epithelial-cell layers separated by a thin extracellular matrix.

Finally, the field-theoretic framework that implies a symmetric coupling between the $Ca^{2+}$ field and the scalar curvature via the action (2), is nontrivial. In general, one expects an asymmetry between the effect of the $Ca^{2+}$ field on the local curvature and the reciprocal effect. In the asymmetric case, the system cannot be described by equations derived from a single action. Instead, one has to write two sets of Langevin-type equations: One for the tissue structure and the second one for the $Ca^{2+}$ activity with an asymmetric coupling between them.

This reciprocal property, whose microscopic origin is unclear[15, 16], is a simplification justified by two arguments. First, the robustness of the morphological changes must include feedback loops[17-23]. Second, analysis of the Granger causality test shows that the information flow from the tissue's shape to the $Ca^{2+}$ activity is of the order of the information flow in the reversed direction, see section E below.

*E. The Granger causality test*

In this section, we show that the information flow from the tissue's shape to the $Ca^{2+}$ activity is of the order of the information flow in the reversed direction. To this end, we employ the Granger causality (GC) test that utilizes time series of sampled data. This test was designed originally to identify the flow of information in the stock markets by studying the time series of stocks[24]. It became widely applied in neuroscience to disentangle the neural network's functional structure[25]. Here, we use it to study causal relations between the shape parameter, $\Lambda(t)$, that characterizes the tissue's shape, and the (spatial) variance of the $Ca^{2+}$ activity, $vr(t)$. We choose the variance rather than the mean because it is more sensitive to morphological changes; the local $Ca^{2+}$ activity is highly localized in the tissue, and these local changes, while may have large effects on the spatial variance might only weakly affect the mean.

For the sake of completeness, we begin by briefly describing the GC test approach. Consider two time series $X_t$ and $Y_t$. By linear regression, one may attempt to predict the value of $X_t$ and $Y_t$ using the information of these series at previous time steps, i.e.,

$$\begin{aligned} X_t &= \sum_{j=1}^{n} a_j X_{t-j} + \sum_{j=1}^{p} b_j Y_{t-j} + \varepsilon_{1t} \\ Y_t &= \sum_{j=1}^{n} c_j Y_{t-j} + \sum_{j=1}^{p} d_j Y_{t-j} + \varepsilon_{2t} \end{aligned} \tag{E.1}$$

where $a_j$ , $b_j$ , $c_j$ and $c_j$ are fixed parameters chosen to minimize the errors of the approximations: $\Gamma_p = \left\langle \varepsilon_{1,t}^2 \right\rangle_t$ and $\Sigma_p = \left\langle \varepsilon_{2,t}^2 \right\rangle_t$, where averaging is over time.

For a fixed value of $n$, $\Sigma_p$ and $\Gamma_p$ are non-decreasing functions of $p$. If $\Gamma_p$ is larger than $\Gamma_0$, the prediction of $X_t$ is improved by knowing the values of $Y_t$ at previous time steps. This property indicates a causal effect of $Y$ on $X$ which can be quantified by the measure of the linear feedback from $Y$ to $X$ [26]:

$$e_{Y\to X} = \ln\left(\frac{\Gamma_0}{\Gamma_p}\right). \tag{E.2}$$

This measure is a non-decreasing function of $p$ that vanishes at $p = 0$. It usually saturates at some value of $p \simeq p_c < n$ for a large enough value of $n$. This saturation value provides an estimate for the typical duration of the causal feedback effect. Similarly, one defines the measure of linear feedback from $X$ on $Y$ as

$$e_{X\to Y} = \ln\left(\frac{\Sigma_0}{\Sigma_p}\right). \tag{E.3}$$

Instantaneous causal relations are generated by additional factors that simultaneously affect both $X$ and $Y$. These are quantified by

$$e_{X\cdot Y} = \ln\left(\frac{\Sigma_p\Gamma_p}{\Sigma_p\Gamma_p - \Upsilon_p^2}\right), \tag{E.4}$$

where $\Upsilon_p = \left\langle \varepsilon_{1,t}\varepsilon_{2,t} \right\rangle_t$ is the cross-correlation of the errors in the two linear regression schemes. The total interdependence or coherence of $X$ and Y is defined by the sum,

$$e_{X,Y} = e_{Y\to X} + e_{X\to Y} + e_{X\cdot Y}, \tag{E.5}$$

also known as the "information measure"[27].

In a finite time series, there are systematic finite-size errors[28]. These can be usually eliminated by constructing random time series of surrogate data and defining the effective measure of the linear feedback to be[29]:

$$e_{X\to Y}^{eff} = e_{X\to Y} - e_{X\to Y}^{sur}, \tag{E.6}$$

where $e_{X\to Y}^{sur}$ represents the feedback measure generated with the surrogate data averaged over many realizations. Similar expressions can be defined for $e_{Y\to X}^{eff}$ and $e_{X\cdot Y}^{eff}$.

An estimate for the error that comes from the finite length of the time series can be deduced by applying the GC test for two independent and stationary random time series, $X_t$ and $Y_t$, containing $N$ time points, assuming that all terms in the series are uncorrelated, with zero mean, $\langle X_t \rangle = \langle Y_t \rangle = 0$, and variances $\langle X_t^2 \rangle = \sigma_X^2$ and $\langle Y_t^2 \rangle = \sigma_Y^2$. Since $a_j$ and $b_j$ are $n+p$ free parameters, one can set them such that $\varepsilon_{1t}$ is identically zero for $n+p$ time steps, say the first ones. We cannot do better because the series are random, uncorrelated, and independent. Taking the ensemble average of $\Gamma_p$ we obtain:

$$\begin{aligned} \langle \Gamma_p \rangle &= \frac{1}{N}\sum_{t=n+p+1}^{N} \langle \varepsilon_{1t}^2 \rangle = \frac{1}{N}\sum_{t=n+p+1}^{N} \left\langle \left( \sum_{j=1}^{n} a_j X_{t-j} + \sum_{j=1}^{p} b_j Y_{t-j} \right)^2 \right\rangle \\ &= \frac{1}{N}\sum_{t=n+p+1}^{N} \left[ \sum_{j=1}^{n} a_j^2 \langle X_{t-j}^2 \rangle + \sum_{j=1}^{p} b_j^2 \langle Y_{t-j}^2 \rangle \right] \\ &= \frac{N-(n+p)}{N}\left[ \sigma_X^2 \sum_{j=1}^{n} a_j^2 + \sigma_Y^2 \sum_{j=1}^{p} b_j^2 \right] \end{aligned} \tag{E.7}$$

On the other hand, for the first $n+p$ terms of the series $\varepsilon_{1t} = 0$. Hence, squaring the upper Eq. (E.1) and averaging over the ensemble of random series, we obtain

$$\sigma_X^2 = \sigma_X^2 \sum_{j=1}^{n} a_j^2 + \sigma_Y^2 \sum_{j=1}^{p} b_j^2 \tag{E.8}$$

From the last two equations, it follows that

$$\Gamma_p = \frac{N-(n+p)}{N}\sigma_X^2. \tag{E.9}$$

A similar calculation gives

$$\Sigma_p = \frac{1}{N}\sum_{t=n+p+1}^{N} \langle \varepsilon_{2t}^2 \rangle = \frac{N-(n+p)}{N}\sigma_Y^2. \tag{E.10}$$

Thus, $e_{X\to Y}^{(\text{random})} = e_{Y\to X}^{(\text{random})}$ with

$$e_{Y\to X}^{(\text{random})} = \log\left(\frac{\Gamma_0}{\Gamma_p}\right) = \log\left(\frac{N-n}{N-n-p}\right) \simeq \frac{p}{N-n}. \tag{E.11}$$

To calculate $e_{X\cdot Y}^{(\text{random})}$, one should estimate the square of the cross-correlation, which is

$$\Upsilon_p^2 \simeq \left\langle \left( \frac{1}{N}\sum_{t=n+p}^{N} \varepsilon_{1t}\varepsilon_{2t} \right)^2 \right\rangle = \frac{N-n-p}{N^2}\left\langle \varepsilon_{1t}^2 \varepsilon_{2t}^2 \right\rangle \qquad \text{(E.12)}$$

where $\left\langle \varepsilon_{1t}^2 \varepsilon_{2t}^2 \right\rangle = \left\langle \varepsilon_{1t}^2 \right\rangle \left\langle \varepsilon_{2t}^2 \right\rangle = \sigma_X^2 \sigma_Y^2$. Thus

$$e_{Y\cdot X}^{(\text{random})} = \log\left( \frac{\Gamma_p \Sigma_p}{\Gamma_p \Sigma_p - \Upsilon_p^2} \right) \qquad \text{(E.13)}$$

$$= \log\left( \frac{\left(1-\frac{n+p}{N}\right)^2}{\left(1-\frac{n+p}{N}\right)^2 - \frac{1}{N}\left(1-\frac{n+p}{N}\right)} \right) \simeq \frac{1}{N-n} + \frac{p}{\left(N-n\right)^2}$$

From the above estimates, it follows that to minimize the finite size errors, one has to set $n$ and $p$ to be the smallest possible values for which $e^{eff}$ still saturates as a function of $p$.

We now describe the procedure we used to generate the surrogate data. The surrogate data are time series that ideally break the causal link but maintain all properties of the underlying statistics. Many ways of generating surrogate data apply to different types of systems[30]. Here, we generate this data by randomizing the phases of the Fourier components of our time series. Thus, the surrogate data has the same power spectrum as the original data. In other words, it maintains the data's actual mean, variance, and correlations. The results presented below are obtained by averaging over 100 realizations of the surrogate data.

Our analysis focuses on the time period far before the transition. However, even in this region, the system cannot be considered stationary due to the poorly understood underlying biological processes. Therefore, the GC test is performed in a window of 300 time steps where underlying changes are small (our previous study[1] shows that the typical relaxation time scale of the tissue is about 100 min). The number of elements in the linear regression is $p \le n = 20$, and we average over different samples as well as over seven different positions of the 300 time steps window that differ by 150 minutes. To ensure small p-values of the unit root test, we work with the difference series of the order parameter and the variance, $\Delta\Lambda(t) = \Lambda(t) - \Lambda(t-1)$, and $\Delta vr(t) = vr(t) - vr(t-1)$, respectively.

It is convenient to present the results in terms of normalized quantities – the measures of the linear feedback normalized by the total interdependence:

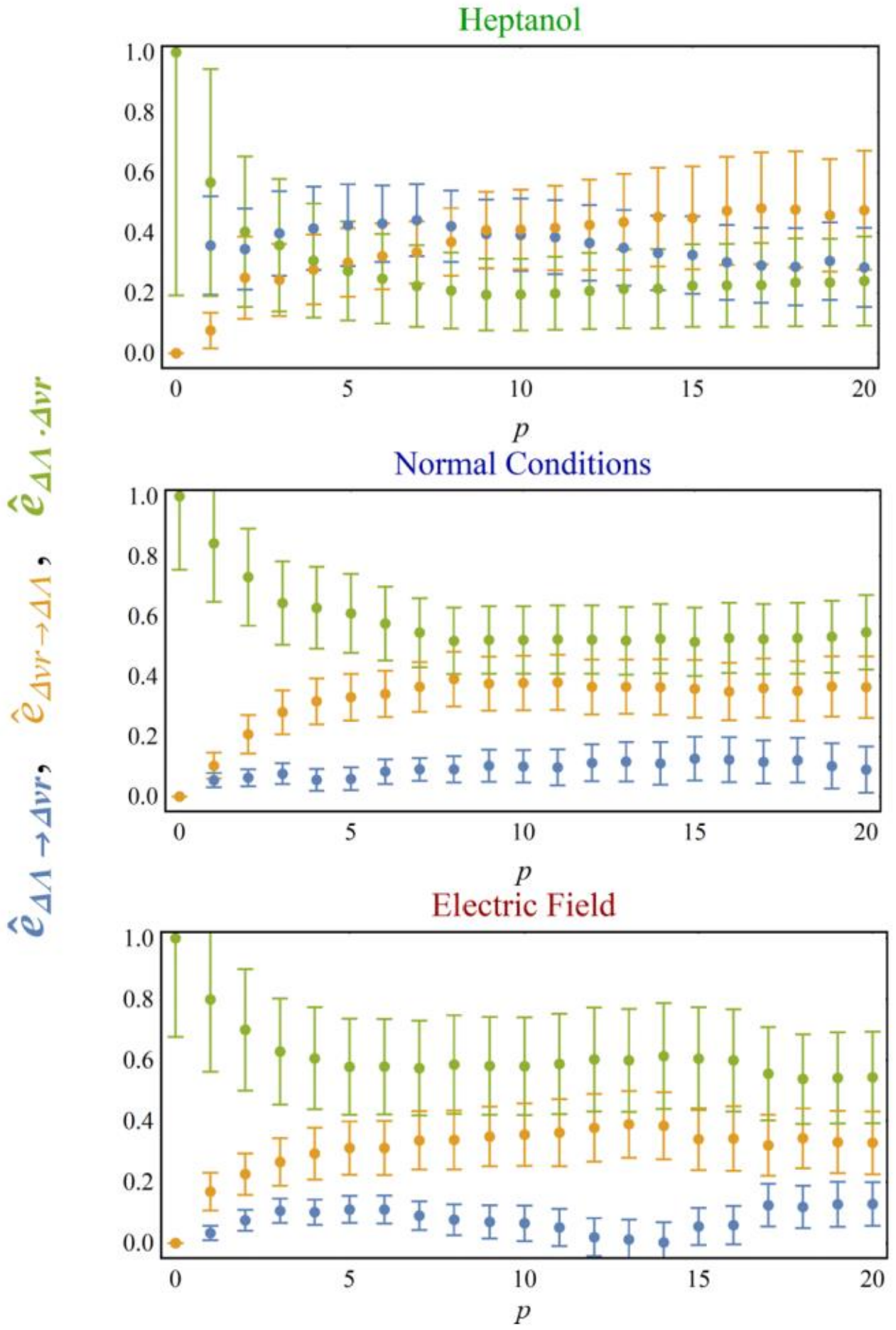

**FIG. S6:** The Granger causality test for the information flow from the $Ca^{2+}$ spatial variance to the order parameter and vice versa. Each panel corresponds to a different control.

$$\begin{aligned} \hat{e}_{\Delta\Lambda\to\Delta vr} &= e^{eff}_{\Delta\Lambda\to vr} \big/ e^{eff}_{\Delta\Lambda,\Delta vr}, \\ \hat{e}_{\Delta vr\to\Delta\Lambda} &= e^{eff}_{\Delta vr\to\Delta\Lambda} \big/ e^{eff}_{\Delta\Lambda,\Delta vr}, \\ \hat{e}_{\Delta\Lambda\cdot\Delta vr} &= e^{eff}_{\Delta\Lambda\cdot\Delta vr} \big/ e^{eff}_{\Delta\Lambda,\Delta vr}. \end{aligned} \qquad \text{(E.14)}$$

In Fig. S6, we present the results for all controls of the experiments described in Fig. 1 of the main text. The blue points represent the flow of information from the tissue's shape to the $Ca^{2+}$ activity distribution, the orange points are associated with information flow in the opposite direction, while the green points describe the instantaneous causality generated by an additional field (or fields) that might affect both $\Delta\Lambda$ and $\Delta vr(t)$ simultaneously.

The normalized measures of the linear feedback (defined by Eq. (E.14)), averaged over the interval $10 \le p \le 20$, are summarized in Table I. With these results, we can quantify the asymmetry in the information flow (from the tissue structure to the calcium activity and vice versa) by the ratio $r_{asy} = \hat{e}_{\Delta\Lambda \to \Delta vr} / \hat{e}_{\Delta vr \to \Delta\Lambda}$ . This ratio is largest for samples subjected to *Heptanol*, $r_{asy} \sim 0.8$ ; smaller for samples in the normal state, $r_{asy} \sim 0.3$; and even smaller for samples subjected to an electric field, $r_{asy} \sim 0.18$. Despite the large error bars, these results support the approximate description given by the field-theoretic Model of Eq. (4), which assumes that information flows reciprocally from the $Ca^{2+}$ field to the structure of the tissue and backward.

TABLE I. The Granger causality test of the shape parameter and the spatial variance of the $Ca^{2+}$ activity normalized by the total interdependence. The left column lists the different controls, and the number of samples appears in parentheses.

| Control | $\hat{e}_{\Delta vr \to \Delta\Lambda}$ | $\hat{e}_{\Delta\Lambda \to \Delta vr}$ | $\hat{e}_{\Delta\Lambda \cdot \Delta vr}$ |
|---|---|---|---|
| Normal (7) | $0.10 \pm 0.07$ | $0.36 \pm 0.09$ | $0.5 \pm 0.1$ |
| Electric Field (6) | $0.06 \pm 0.06$ | $0.35 \pm 0.10$ | $0.6 \pm 0.2$ |
| Heptanol (6) | $0.3 \pm 0.1$ | $0.4 \pm 0.2$ | $0.2 \pm 0.1$ |

In the normal case and for samples subjected to an electric field, the asymmetry in the information flow appears to be significant. However, one should take into account the inherent limitation of the GC test for complex systems, which might mask the true nature of the information flow. To explain this point, consider, e.g., the system depicted in Fig. S7. This system contains three fields, $X, Y$ and $Z$ , but only $X$ and $Y$ can be accessed and measured. Here, $Z$ is affected by $Y$ while simultaneously affecting both $X$ and $Y$ . Thus, there is an indirect information flow from $Y$ to $X$ , but it is hidden by the instantaneous causality generated by the $Z$ -field. Thus, a small value of $r_{asy}$ does not necessarily mean there is no information flow from the tissue's geometrical structure to the calcium activity.

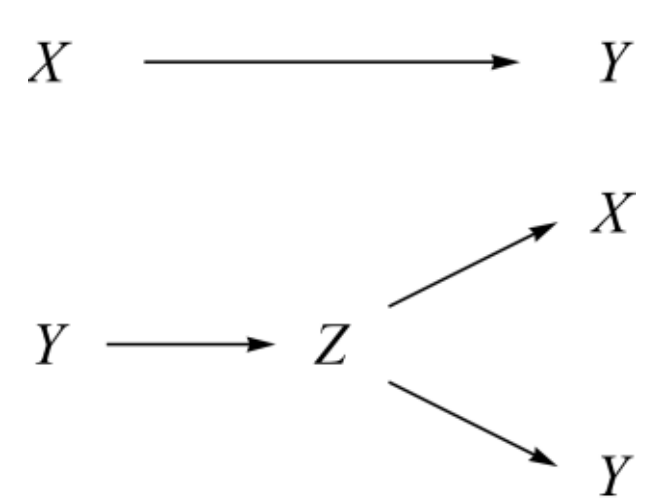

**FIG. S7:** An example of a system where the instantaneous causal effect due to the field $Z$ hides the causal effect of $Y$ on $X$ .

*F. The morphological potential,* $V(q)$

The potential $V(q)$ is defined by the integral (6) over all configurations of the $Ca^{2+}$ field, assuming the tissue shape is given by Eq. (5). This integral is evaluated approximately as explained in this section.

First, one should notice that spatial morphological changes in the tissue are slow, while underlying the $Ca^{2+}$ field, are both slow and fast degrees of freedom. However, since the scalar curvature is a slowly varying field, the integral over space in $S_{\text{int}}$ picks up only the slow components of the $\phi$ field. The fast components of $\phi$ can be integrated out without significant effect on the value of $V(q)$ . Hence, we choose to represent the slow component of the $Ca^{2+}$ field by:

$$\phi_{slow}(\theta) = \phi_0 f(q)\left[1 + v\left(3\cos^2\theta - 1\right)\right], \qquad \text{(F.1)}$$

with $-0.5 < v < 1$ and $\phi_0 > 0$ to secure positivity. The parameter $v$ dictates the form of distribution of the slow component of the calcium field over the tissue, while $\phi_0$ determines the amplitude of the field. The function $f(q)$, which is positive and equals one for a sphere, $f(0) = 1$, is chosen to ensure that the total $Ca^{2+}$ activity over the *Hydra*'s tissue reflects the intracellular processes rather than changes in the tissue's area. The latters are significant along the regeneration process, while the number of cells remains approximately constant throughout the morphological transition[31].

To obtain $V(q)$, one first finds the values of $\phi_0^{\min}(q)$ and $v^{\min}(q)$ that minimize the action (4) for a given value of $q$ , and then expands the action to quadratic order around this point:

$$S \simeq S\left(\phi_0^{\min}, v^{\min}\right) + \frac{1}{2}\left(\phi_0 - \phi_0^{\min}, v - v^{\min}\right) M \begin{pmatrix} \phi_0 - \phi_0^{\min} \\ v - v^{\min} \end{pmatrix}, \qquad \text{(F.2)}$$

where the elements of the matrix $M$ are the second derivatives of $S$ evaluated at the minimal point. The integration over $\phi_0$ and $v$ with the proper Jacobian yields the potential

$$V(q) \simeq S\left(\phi_0^{\min}, v^{\min}\right), \qquad \text{(F.3)}$$

up to a logarithmic correction that comes from the fluctuations of the $\phi_{\text{slow}}$ near the minimum. The latter contribution is neglected since the sharpness of the morphological transition suggests that the barrier between the two minima of $V(q)$ is high, implying that the fluctuations around the minimum are costly. In what follows, we describe the calculation of $\phi_0^{\min}(q)$ and $v^{\min}(q)$ in detail.

A single parameter $q$ parametrizes the tissue shape. Hence one can fix either the volume enclosed by the tissue or its area (or some combination of the two). We fix the volume because of the incompressible nature of the fluid in the internal cavity, which is only modulated due to osmotic pressure gradients that do not seem to affect the phase transition directly. On the other hand, the flexibility of the tissue allows for large changes in its area, which in principle, is only weakly constrained. The volume is given by

$$Vol = \frac{4\pi R_0^3}{3}\left[1+q^2\left(\frac{12}{5}+\frac{16}{35}q\right)\right], \tag{F.4}$$

and fixing it to be the volume of a unit sphere (without loss of generality) yields:

$$R_0(q) = \left[1+q^2\left(\frac{12}{5}+\frac{16}{35}q\right)\right]^{-1/3}. \tag{F.5}$$

We turn now to calculate the extrinsic and intrinsic curvatures of the surface. The tangent vectors to the surface are:

$$\begin{aligned} \boldsymbol{r}_\theta &= \frac{\partial}{\partial\theta}R(\theta)(\sin\theta\cos\varphi,\sin\theta\sin\varphi,\cos\theta) \\ \boldsymbol{r}_\varphi &= \frac{\partial}{\partial\varphi}R(\theta)(\sin\theta\cos\varphi,\sin\theta\sin\varphi,\cos\theta) \end{aligned} \tag{F.6}$$

Hence the metric tensor is diagonal, and its components are:

$$g_{\theta\theta} = \boldsymbol{r}_\theta\cdot\boldsymbol{r}_\theta\ ,\ g_{\varphi\varphi} = \boldsymbol{r}_\varphi\cdot\boldsymbol{r}_\varphi\ ,\ g_{\theta\varphi} = 0\ . \tag{F.7}$$

The square of the Jacobian is given by

$$g(q) = \det g_{\mu\nu}\ , \tag{F.8}$$

and a unit vector normal to the surface and pointing outwards is :

$$\hat{\boldsymbol{n}} = \frac{\boldsymbol{r}_\theta\times\boldsymbol{r}_\varphi}{\left|\boldsymbol{r}_\theta\times\boldsymbol{r}_\varphi\right|} = \frac{\boldsymbol{r}_\theta\times\boldsymbol{r}_\varphi}{\sqrt{g(q)}}\ . \tag{F.9}$$

With the above definitions, the mean curvature is given by

$$\mathcal{K} = g^{\mu\nu}K_{\mu\nu} = \frac{K_{\theta\theta}}{g_{\theta\theta}} + \frac{K_{\varphi\varphi}}{g_{\varphi\varphi}} \tag{F.10}$$

where

$$K_{\mu\nu} = -\boldsymbol{n}\cdot\partial_\mu\partial_\nu\boldsymbol{r} \tag{F.11}$$

is the extrinsic curvature tensor. The above expression is used for calculating the elastic part of the action given by Eq. (3).

To calculate the part of the action that describes the $Ca^{2+}$ fluctuations, we impose the condition that the average $Ca^{2+}$ activity within individual tissue cells does not change by their contraction or extension. Therefore, the function $f(q)$ in F1, is set to be inversely proportional to the Jacobian of the transformation from the sphere to an elongated spheroid, i.e.

$$f(q) = \sqrt{\frac{g(0)}{g(q)}}\ . \tag{F.12}$$

This condition ensures that the total integral of $\phi_{slow}(\theta)$ over the surface is fixed to be $4\pi\phi_0$. Notice, however, that the amplitude $\phi_0$ can fluctuate.

The potential of the $Ca^{2+}$ field is chosen to be the one found to approximately fit the experimental data (in the normal case), see Ref.[1]:

$$U(\phi) = \frac{A}{\phi} + \left(1-\Delta\log\cosh\left[(\phi-\nu)/\Delta\right]\right)^2 + \mu\phi, \tag{F.13}$$

with $A = 0.005$, $\Delta = 0.32$, $\nu = 0.24$ and $\mu = 1.3$.

Finally, to obtain the coupling action of Eq. (2), we extract the scalar curvature from the contraction of the Gauss-Codazzi-Mainardi equations, which relate the intrinsic and extrinsic curvatures:

$$\mathcal{R} = \mathcal{K}^2 - K^{\mu\nu}K_{\mu\nu} = \mathcal{K}^2 - \left(\frac{K_{\theta\theta}}{g_{\theta\theta}}\right)^2 - \left(\frac{K_{\varphi\varphi}}{g_{\varphi\varphi}}\right)^2. \tag{F.14}$$

The total action obtained from the above equations, after integration over $\theta$, depends on the following parameters: $B$ (the bending modulus), $\mathcal{K}_0$ (the spontaneous tissue's curvature), and $\eta$ (the coupling strength). It also depends on the variables, $\nu$, $\phi_0$ and $q$. The precise values of the parameters, $B$ and $\mathcal{K}_0$ (as long as they are positive) dictate the value of the critical coupling parameter $\eta_c$, required for the transition.

To obtain Fig. 2, we set $B = 0.1$, and choose $\mathcal{K}_0 = 2$; namely, the spontaneous curvature of the tissue is set to be the curvature of the initial sphere. We have checked that the qualitative behavior of the system is insensitive to this particular choice of parameters.

The stiffness parameter determines the value of the coupling constant at the transition $\eta = \eta_c$, as well as the point $q_c$ where the transition takes place. As $D$ increases, $q_c$ decreases while $\eta_c$ increases. When $D \ll 1$ the transition occurs at $q_c \simeq 1$, i.e., close to the maximal possible value of the shape parameter, at which the surface deforms into two closed surfaces touching at a point. To obtain the potentials shown in Fig. 2 we use $D=4$ for which $q_c \simeq 0.5$, and coupling strengths, $\eta =$ 2.1, 2.485, and 2.8, corresponding to situations that describe the behavior prior, at, and after the transition.

Now, for each value of $q$ we calculate the action $S(q,v,\phi_0)$ over a grid of $100\times 100$ points in the space of $(v,\phi_0)$. Next, we construct an interpolating function and identify the minimum point, $\left(v^{\min}(q),\phi_0^{\min}(q)\right)$ which determines the value of the potential at $q$ by Eq. (F.3). Finally, we comment that unless $D$ is very large for each fixed value of $q$,the action $S(q,v,\phi_0)$ has a single minimum.

*G. The instanton formula, Eq. (8)*

The instanton formula describing the transition in a double-well potential, $V(q)$, is well known; see, for example, Refs.[31, 32]. Here, for completeness, we provide its derivation and the approximate formula given by Eq. (8) that holds when the noise is moderately strong. Our starting point is the Langevin equation:

$$\frac{dq}{dt} = f(q)+\xi(t), \tag{G.1}$$

with the force

$$f(q) = -\frac{d}{dq}V(q), \tag{G.2}$$

and a $\delta$-correlated Gaussian noise, $\xi(t)$, of zero mean. For simplicity, we choose the double-well potential to be symmetric with minima at $q=\pm 1$, thus

$$V(q) = \lambda\left(1-q^2\right)^2, \tag{G.3}$$

where $\lambda$ controls the height of the barrier between the two minima.

Employing the Martin-Siggia-Rose approach[33] we construct the field-theoretic description of the transition. Namely, we start by calculating the "partition function":

$$\begin{aligned} Z &= \left\langle \prod_t \delta\left(\frac{dq(t)}{dt} - f\left[q(t)\right]-\xi(t)\right)\right\rangle \qquad \text{(G.4)} \\ &= \int D\eta(t)\left\langle \exp\left\{-i\int dt'\eta(t')\left[\frac{dq(t')}{dt} - f\left[q(t')\right]-\xi(t')\right]\right\}\right\rangle. \end{aligned}$$

Averaging over the Gaussian noise, taking into account that $\left\langle \xi(t)\xi(t')\right\rangle = \sigma^2\delta(t-t')$ yields:

$$\begin{aligned} Z = \int D\eta(t)\exp\Bigg(\int dt'\Bigg\{-i\eta(t')\left[\frac{dq(t')}{dt} - f\left[q(t')\right]\right] \\ -\frac{\sigma^2\eta^2(t')}{2}\Bigg\}\Bigg) \end{aligned} \tag{G.5}$$

Integrating over $\eta(t)$, in the Itô sense, we obtain:

$$Z = \exp\left(-\frac{S}{2\sigma^2}\right) \tag{G.6}$$

with the action

$$S = \int dt'\left(\frac{dq(t')}{dt}\right)^2 + f^2\left[q(t')\right] - \sigma^2 V''\left[q(t')\right]. \tag{G.7}$$

When the noise is weak compared to the typical value of the action, the saddle point approximation dictates the dynamics. In this case, the variation of $S$ gives:

$$2\frac{d^2q}{dt^2} = \frac{\partial}{\partial q}\left[f^2(q) - \sigma^2 V''(q)\right]. \tag{G.8}$$

Multiplying this equation by $\partial q/\partial t$ allows us to integrate it. The resulting equation is:

$$\left(\frac{dq}{dt}\right)^2 + P(q) = E, \tag{G.9}$$

where $E$ is the constant of integration, while

$$P(q) = -\left[V'(q)\right]^2 + \sigma^2 V''(q) \tag{G.10}$$

plays a role similar to potential energy. The instanton is obtained from the solution of Eq. (G.9) when setting $E$ to the maximal value of this potential, i.e.

$$E = P(q_{\max}) \text{ with } q_{\max} = \sqrt{\frac{2}{3}+\frac{1}{3}\sqrt{1+\frac{9\sigma^2}{4\lambda}}}\,. \tag{G.11}$$

Thus, from (G.9-11), we obtain:

$$\pm\int_{-q_{\max}}^{q}\frac{dq'}{\sqrt{P(q_{\max})-P(q')}} = t-t_c, \tag{G.12}$$

where $t_c$ is the integration constant. The integral on the left-hand side of this equation can be expressed in terms of the inverse of the Jacoby amplitude function. Typical solutions of (G.12) for various values of the noise are depicted in Fig. S8. From this figure, it follows that for not too weak noise, the solution can be approximated by:

$$q(t) = q_0 \tanh(\beta t), \tag{G.13}$$

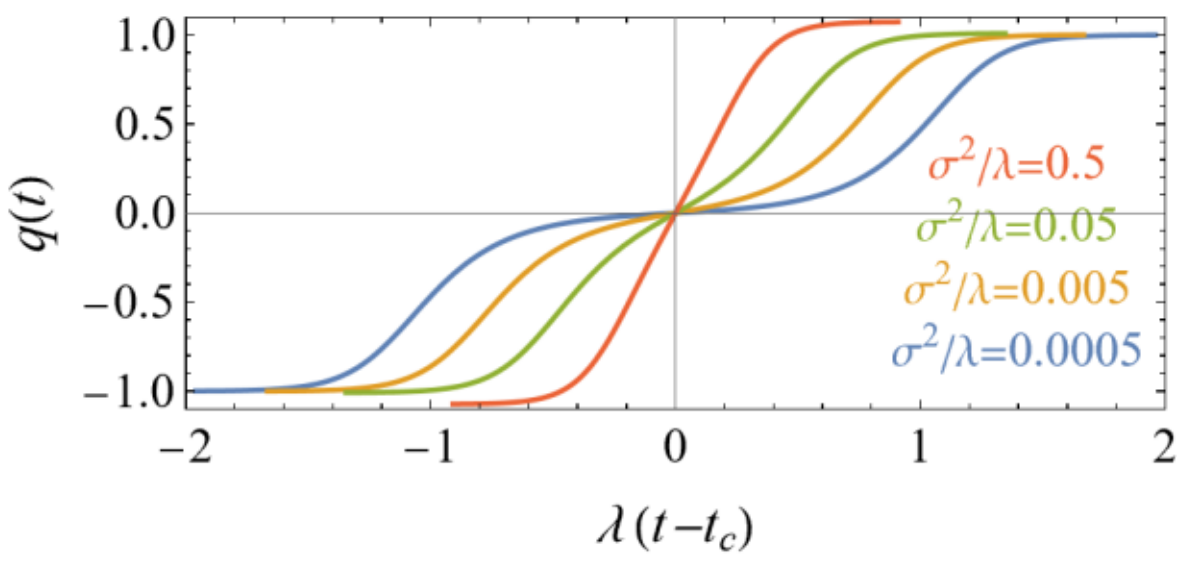

**FIG. S8:** Solutions of the instanton equation (G.8) for various values of the ratio of the noise strength to the barrier height $\sigma^2/\lambda$. When $\sigma^2/\lambda \sim 1$ the solution can be approximately described by (G.13). At much higher values, the instanton solution becomes meaningless because fluctuations around the saddle point are too strong and the system passes rapidly between the two minima of the double well.

where $q_0$ and $\beta$ are variational parameters. The validity of this approximation has also been checked by numerical solutions of the Langevin equation (G.1).

Substituting (G.13) in the action (G.7) and integrating from $t_i$ to $t_f$, assuming $t_f = -t_i \to \infty$, we obtain an action that depends on two parameters, $q_0$ and $\beta$. Minimizing, first, with respect to $q_0$, we get $q_0 = q_{\max}$. Next, minimizing with respect to $\beta$, we arrive at

$$\beta = 2\sqrt{\frac{11}{5}}\sqrt{\lambda}\sigma\sqrt{1-\frac{20\lambda}{99\sigma^2}\left(1-\frac{14}{5}\sqrt{1+\frac{9\sigma^2}{4\lambda}}\right)} \quad . \qquad \text{(G.14)}$$

For large enough $\sigma^2/\lambda$, the above expression reduces to:

$$\beta \simeq 2\sqrt{\frac{11}{5}}\sqrt{\lambda}\sigma = 2.97\sqrt{\lambda}\sigma \,. \qquad \text{(G.15)}$$

From here, we obtain that the instanton solution describing the transition between the minima, when the noise is moderately strong, is approximately given by Eq. (8).

### *H. The approximate formula for* $M_2$

We assume the *Hydra*'s surface to be close to an ellipsoid,

$$\alpha^2 x^2 + y^2 + z^2 = R^2 \qquad \text{(H.1)}$$

whose symmetry axis is parallel to the projection imaging plane, as shown in Fig. S9. Here $\alpha$ is the aspect ratio of the principal axes of the projected ellipse that can be expressed as $\alpha = 1-e^2$ where $0<e<1$ is the eccentricity of the projected ellipse on the $xy$ plane, and $R$ is its semiminor axis. An arbitrary vector on the lower half of the spheroid surface is given by:

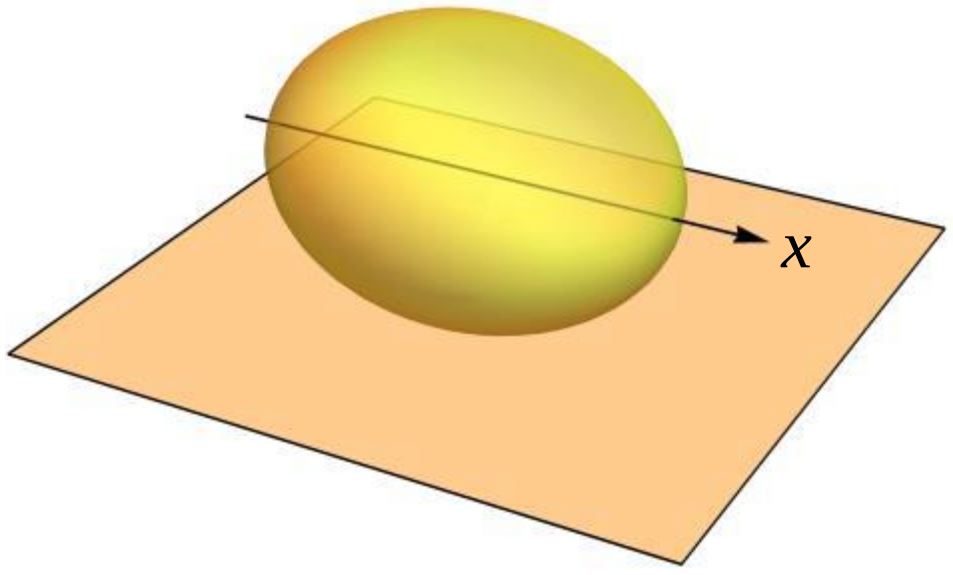

**FIG. S9:** The *Hydra* surface approximated by a spheroid whose axis is parallel to the projection imaging plane.

$$\boldsymbol{r}(x,y) = \left(x, y, -\sqrt{R^2-\alpha^2x^2-y^2}\right). \qquad \text{(H.2)}$$

Thus, two vectors that are tangent to the surface are:

$$\begin{aligned}\boldsymbol{r}_x &= \frac{\partial \boldsymbol{r}}{\partial x} = \left(1, 0, \frac{\alpha^2 x}{\sqrt{R^2-\alpha^2x^2-y^2}}\right),\\ \boldsymbol{r}_y &= \frac{\partial \boldsymbol{r}}{\partial y} = \left(0, 1, \frac{y}{\sqrt{R^2-\alpha^2x^2-y^2}}\right),\end{aligned} \qquad \text{(H.3)}$$

and an infinitesimal area element on the spheroid is given by:

$$, dA = \left|\boldsymbol{r}_x \times \boldsymbol{r}_y\right| dxdy = \Omega(x,y)\,dxdy\,, \qquad \text{(H.4)}$$

where

$$\Omega(x,y) = \sqrt{\frac{R^2-\alpha^2\left(1-\alpha^2\right)x^2}{R^2-\alpha^2x^2-y^2}}\,. \qquad \text{(H.5)}$$

Thus if $\phi(\boldsymbol{r})$ represents the $Ca^{2+}$ signal on the *Hydra*'s surface, the projected signal is given by

$$\phi_p(x,y) = \phi(\boldsymbol{r})\Omega(x,y)\,. \qquad \text{(H.6)}$$

In this formula, the square root divergence of $\phi_p(x,y)$ near the edge of the projected image is due to the assumption of zero tissue thickness. This singularity disappears when taking into account the finite thickness of the tissue. However, since the singularity is integrable, one can neglect finite thickness effects in the leading approximation.

The average signal can be calculated from the projected signal by

$$\bar{\phi} = \frac{1}{A}\oint\!\!\oint_{\text{spheroid}} dA\phi(\boldsymbol{r}) = \frac{2}{A}\iint_{\substack{\text{projected}\\ \text{spheroid}}} dxdy\phi_p(x,y)\,, \qquad \text{(H.7)}$$

where $A$ is the total area of the spheroid. The factor of two on the right-hand side of the equation takes into account that the projected signal comes essentially from the lower part of the spheroid. It is assumed that the statistical behavior of the signal in the upper half part of the tissue is the same as in the lower part.

The total area of the spheroid is given by

$$A = 2 \iint_{\alpha^2 x^2 + y^2 < R^2} dxdy\Omega(x,y) = 2\pi R^2 \left[1 + \frac{\cos^{-1}\alpha}{\alpha\sqrt{1-\alpha^2}}\right]. \quad \text{(H.8)}$$

To calculate the second moment of the $Ca^{2+}$ activity, we need the following integral

$$I_2 = \oint\!\!\!\oint dA|\boldsymbol{r}|\phi(\boldsymbol{r}) = 2 \iint_{x^2+\alpha^2 y^2 < R^2} dxdy\left(x^2+y^2+z^2\right)\phi_p(x,y)$$
$$= 2 \iint_{x^2+\alpha^2 y^2 < R^2} dxdy\left[R^2 + \left(1-\alpha^2\right)x^2\right]\phi_p(x,y), \quad \text{(H.9)}$$

where we have used the relation: $z^2 = R^2 - \alpha^2 x^2 - y^2$.

From the above integral, we subtract the second moment that one would have obtained if the $Ca^{2+}$ activity was uniform over the spheroid, $\bar{I}_2$. This quantity is obtained from the above integral by replacing $\phi_p(x,y)$ with a constant average activity on the surface $\bar{\phi}$. Thus,

$$\bar{I}_2 = \oint\!\!\!\oint dA|\boldsymbol{r}|^2\bar{\phi} = 2\bar{\phi} \iint_{x^2+\alpha^2 y^2 < R^2} dxdy\left(x^2+y^2+z^2\right)\Omega(x,y)$$
$$= 2\bar{\phi} \iint_{\alpha^2 x^2 + y^2 < R^2} dxdy\left[R^2 + x^2\left(1-\alpha^2\right)\right]\sqrt{\frac{R^2-\alpha^2\left(1-\alpha^2\right)x^2}{R^2-\alpha^2 x^2 - y^2}}$$
$$= \frac{\pi\bar{\phi}R^4}{2}\left[2 + \frac{1}{\alpha^2} + \frac{1+4\alpha^2}{\alpha^3\sqrt{1-\alpha^2}}\cos^{-1}\alpha\right]. \quad \text{(H.10)}$$

By substituting the above results in Eq. (9) we obtain the approximate formula for the normalized second moment of the calcium activity expressed in terms of the projected signal:

$$M_2 = \frac{I_2}{\bar{I}_2} - 1$$
$$= \frac{\iint_{\alpha^2 x^2 + y^2 < R^2} dxdy\left[R^2 + \left(1-\alpha^2\right)x^2\right]\phi_p(x,y)}{\frac{\pi\bar{\phi}R^4}{4}\left[2 + \frac{1}{\alpha^2} + \frac{1+4\alpha^2}{\alpha^3\sqrt{1-\alpha^2}}\cos^{-1}\alpha\right]} - 1. \quad \text{(H.11)}$$

From the above formula it follows that on a sphere, $M_2$ vanishes; Thus, one can derive an approximate formula for $M_2$ when $\alpha \to 1$. In this limit $I_2$ reduces to

$$I_2 = 2 \iint_{\alpha^2 x^2 + y^2 < R^2} dxdy\left[R^2 + \left(1-\alpha^2\right)x^2\right]\phi_p(x,y)$$
$$= A\left[\bar{\phi}R^2 + \left(1-\alpha^2\right)R^2\Delta\phi\right], \quad \text{(H.12)}$$

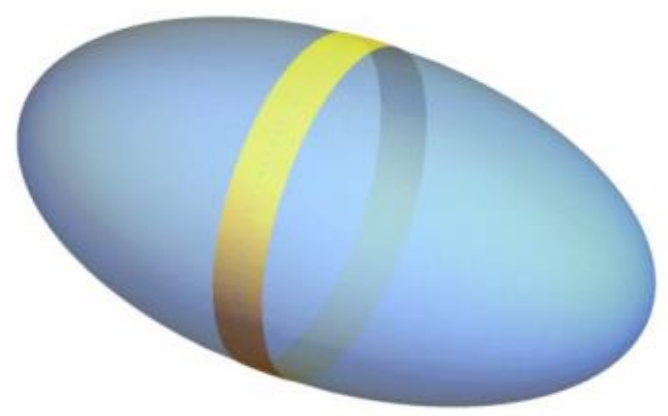

**FIG. S10:** The calcium activity pattern used for calculating a lower bound on $M_2$: The activity is concentrated on the equator of the spheroid.

where

$$\Delta\phi = \frac{2}{A} \iint_{\alpha^2 x^2 + y^2 < R^2} dxdy \frac{x^2}{R^2}\phi_p(x,y), \quad \text{(H.13)}$$

$$A \simeq 4\pi R^2\left[1 + \frac{2}{3}(1-\alpha)\right], \quad \text{(H.14)}$$

and

$$\bar{I}_2 \simeq 4\pi R^4\bar{\phi}\left[1 + \frac{4}{3}(1-\alpha)\right]. \quad \text{(H.15)}$$

Substituting these results in (H.11) and using $\alpha = 1 - e^2$ we obtain

$$M_2 \simeq 2\left(\frac{\Delta\phi}{\bar{\phi}} - \frac{1}{3}\right)e^2. \quad \text{(H.16)}$$

Thus $M_2 \xrightarrow[e\to 0]{} 0$ regardless of the $e$ dependence of $\Delta\phi/\bar{\phi}$ (taking the reasonable assumption that it is not singular when $e \to 0$). The observation that $M_2$ is negative implies that $\Delta\phi < \bar{\phi}/3$, where $\bar{\phi}/3$ is the value obtained for a uniform distribution of the $Ca^{2+}$ activity. The approximate parabolic shape of the scatter plot of $M_2 - e^2$ near $e = 0$ implies that $\Delta\phi/\bar{\phi} \simeq (1/3) - ce^2$, where $c$ is a constant. Thus, the distribution of the $Ca^{2+}$ activity changes with the shape of the tissue such that it is more concentrated near regions of small curvature (i.e., the equator of the spheroid).

To appreciate the magnitude of $M_2$, let us calculate a lower bound on its value. This lower bound is obtained when assuming the $Ca^{2+}$ activity to be concentrated on a narrow stripe of width $w \to 0$ located on the spheroid equator, as demonstrated in Fig. S10, and taking the limit of this width to zero.

Assuming a uniform activity $\phi_*$ within the stripe, we obtain:

$$\bar{\phi} = \frac{2\pi R w\phi_*}{A}, \quad \text{(H.17)}$$

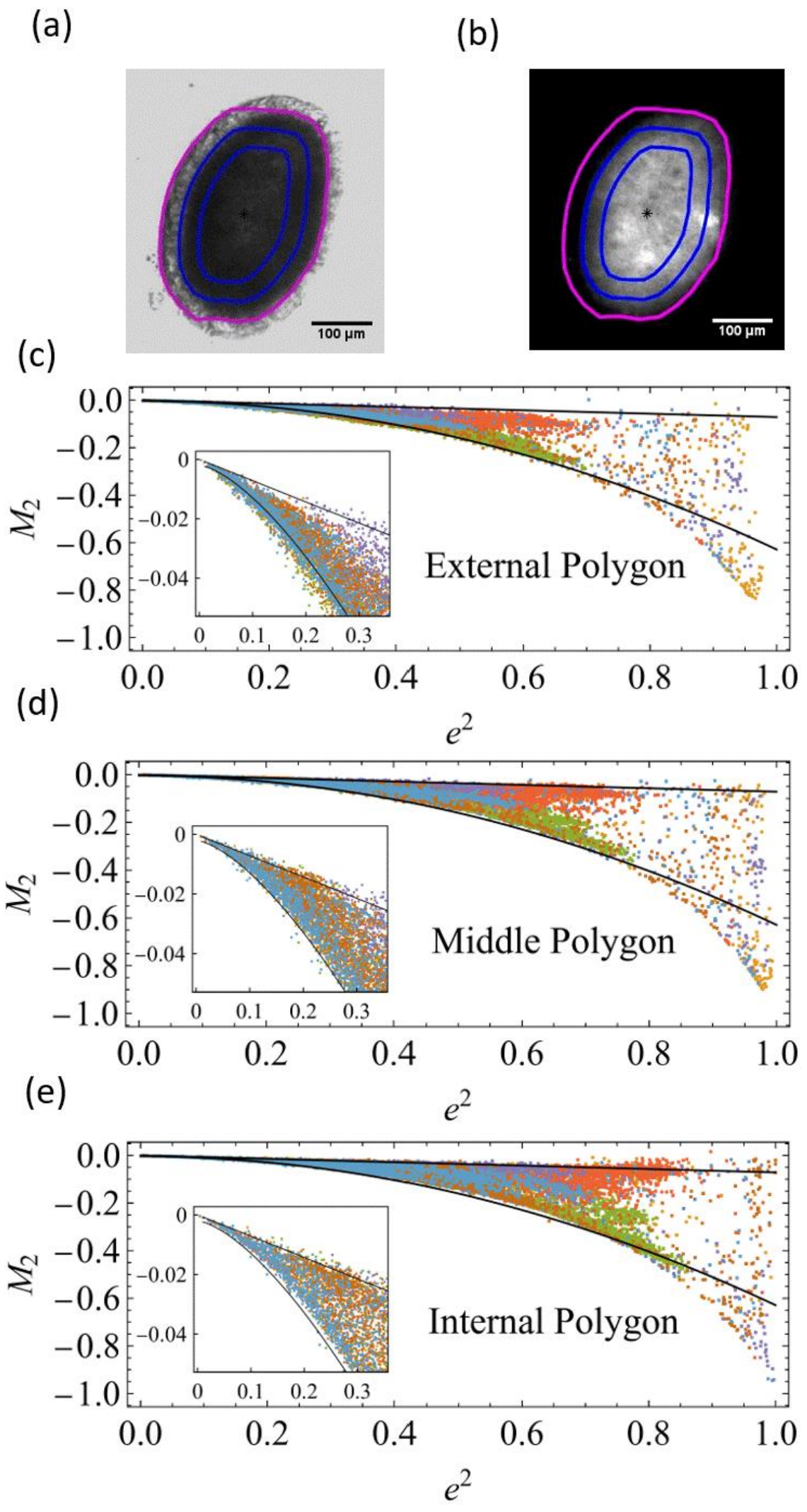

**FIG. S11:** A bright-field $(a)$ and the corresponding fluorescence image $(b)$ of the *Hydra*'s tissue with three polygons of different sizes. The lower panels $(c)$, $(d)$ and $(e)$ show the second moment of the $Ca^{2+}$ activity, $M_2$, calculated for reduced polygons that diminish the optical projection effects. The data is for the normal case. The black lines that serve as a guide for the eye are the same in all panels. The upper line is the straight line: $M_2 = -e^2/14$, while the lower line is the inverted parabola, $M_2 = -0.004 - e^4/1.6$

and

$$I_2 = 2\pi R^3 w \phi_* . \qquad \text{(H.18)}$$

Using these formulas to calculate $\bar{I}_2$ by (H.11), and $M_2$ by (H.12) and obtain:

$$M_2 > M_2^{LB} = \frac{\alpha\sqrt{1-\alpha^2}\left(2\alpha^2-1\right) - \cos^{-1}\alpha}{\alpha\sqrt{1-\alpha^2}\left(2\alpha^2+1\right) + \left(1+4\alpha^2\right)\cos^{-1}\alpha}$$

(H.19)

This lower bound, which is drawn by the dotted lines in Figs. 3c, 3d and 3e, has the following asymptotic behavior:

$$M_2^{LB} = \begin{cases} -2e^2/3 & e \ll 1 \\ -1 + 4\left(1-e^2\right)^2 & e \sim 1 \end{cases} . \qquad \text{(H.20)}$$

We turn now to discuss possible systematic errors in the estimation of $M_2$. One possibility is an enhancement of the $Ca^{2+}$ activity due to its contact with the holding setup. The microscope objective is located below the setup (an inverted microscope), and the fluorescence signal is collected mainly from the lower part of the tissue, which may be in contact with the bottom of the holding setup. If this contact generates excess activity, it will result in a negative value that does not represent direct correlations between the $Ca^{2+}$ activity and the local curvature.

However, our studies of the fluorescence signal in nearly spherical tissues (see SI of Ref.[1]) do not indicate a significant enhancement effect at a possible contact region. Moreover, experimenting with a soft gel, that alters the contact interaction, also does not show a significant change in the radial profile of the fluorescence signal (see SI of Ref.[1]).

Two additional sources of errors are; optical projection effects and the finite thickness of the tissue. In Ref.[1], it was shown that these errors are considerably reduced by limiting the analysis to a smaller internal part of the tissue. In Figs. S11a and S11b, we depict an example of the projected bright-field and corresponding fluorescence image of the *Hydra*'s tissue, together with three different polygons that are parallel to the tissue's contour. The external one is obtained from the bright field image of the tissue and follows the actual boundary of the projected image. The middle polygon is obtained by reducing the outer polygon evenly (parallel to the original polygon) by 20%. The inner polygon is obtained by further reduction of an additional 20% compared to the middle polygon. With this choice, the three polygons share the same aspect ratio. In the lower three panels of Fig. S11 we present the results for $M_2$, calculated for these three converging polygons. The black lines in all three figures are the same and only serve as a guide to the eye. This figure shows that the projection effects lead to minor changes in the evaluation of $M_2$. The largest difference is between the results of the outer and the middle polygons. The reason is that the outer polygon includes the outer cellular layer of the tissue, which is dark due to a reduction in the fluorescence level at the edges due to geometrical scattering effects. The scatter plots of $M_2$ shown in Fig. 3 of the main text are calculated using the middle polygon.

*I. References*

[1] O. Agam and E. Braun, BioRxiv: https://doi.org/10.1101/2021.11.01.466811; arxiv: https://doi.org/10.48550/arXiv.2303.02671 (2023).

[2] E. Braun and H. Ori, Biophysical Journal **117**, 1514 (2019).

[3] S. G. Cummings and H. R. Bode, Wilhelm Roux's archives of developmental biology **194**, 79 (1984).

[4] A. Gierer and H. Meinhardt, Kybernetik **12**, 30 (1972).

[5] H. D. Park, A. B. Ortmeyer, and D. P. Blankenbaker, (1970).

[6] C. Futterer, C. Colombo, F. Julicher, et al., EPL **64**, 137 (2003).

[7] J. Soriano, C. Colombo, and A. Ott, PHYSICAL REVIEW LETTERS **97**, 258102 (2006).

[8] J. Soriano, S. Rudiger, P. Pullarkat, et al., Biophys J **96**, 1649 (2009).

[9] A. Livshits, L. Shani-Zerbib, Y. Maroudas-Sacks, et al., Cell Reports **18**, 1410 (2017).

[10] Y. Maroudas-Sacks, L. Garion, L. Shani-Zerbib, et al., Nature Physics **17**, 251 (2021).

[11] L. Shani-Zerbib, L. Garion, Y. Maroudas-Sacks, et al., Genes **13**, 360 (2022).

[12] A. Hernandez, M. F. Staddon, M. Moshe, et al., arXiv:2303.06224v1 (2023).

[13] N. Murisic, V. Hakim, IoannisÂ G. Kevrekidis, et al., Biophysical Journal **109**, 154 (2015).

[14] A. Gierer, S. Berking, H. Bode, et al., Nature/New Biology, 98 (1972).

[15] E. Farge, Current Biology **13**, 1365 (2003).

[16] C. Guillot and T. Lecuit, Science **340**, 1185 (2013).

[17] L. V. Beloussov, S. V. Saveliev, I. I. Naumidi, et al., in *International Review of Cytology* (Academic Press, 1994), Vol. 150, p. 1.

[18] E. Braun and K. Keren, BioEssays **40**, 1700204 (2018).

[19] E. Hannezo and C.-P. Heisenberg, Cell **178**, 12 (2019).

[20] J. Howard, S. W. Grill, and J. S. Bois, Nature Review Molecular Cell Biology **12**, 392 (2011).

[21] T. Mammoto and D. E. Ingber, Development **137**, 1407 (2010).

[22] A. Zakharov and K. Dasbiswas, The European Physical Journal E **44**, 82 (2021).

[23] A. Zakharov and K. Dasbiswas, Soft Matter **17**, 4738 (2021).

[24] C. W. J. Granger, Econometrica **37**, 424 (1969).

[25] S. Ali and B. F. Emily, Annual Review of Statistics and Its Application **9**, 289 (2022).

[26] W. S. W. C. F. p. d. J. William, Journal of the American Statistical Association **77**, 316 (1982).

[27] I. M. Gel'fand and A. M. Yaglon, *Calculation of the amount of information about a random function contained in another such function contained in another such function*, 1959).

[28] F. Abdul Razak and H. J. Jensen, PLOS ONE **9**, e99462 (2014).

[29] P. Jizba, H. .Lavička, and Z. Tabachová, Entropy **24**, 855 (2022).

[30] T. Schreiber and A. Schmitz, Physica D: Nonlinear Phenomena **142**, 346 (2000).

[31] L. C. H. Katharine and R. John, The Journal of Chemical Physics **75**, 976 (1981).

[32] U. Weiss, Physical Review A **25**, 2444 (1982).

[33] P. C. Martin, E. D. Siggia, and H. A. Rose, Physical Review A **8**, 423 (1973).